\documentclass[fleqn,usenatbib]{mnras}

\usepackage{newtxtext,newtxmath}

\usepackage[T1]{fontenc}

\DeclareRobustCommand{\VAN}[3]{#2}
\let\VANthebibliography\thebibliography
\def\thebibliography{\DeclareRobustCommand{\VAN}[3]{##3}\VANthebibliography}

\usepackage{graphicx}	
\usepackage{amsmath}	
\usepackage{natbib}
\usepackage{pdflscape}
\usepackage{threeparttable}

\newcommand{\Gaia}{\textit{Gaia}}
\newcommand{\degree}{$^{\circ}$}
\newcommand{\kmsec}{\mbox{km\,s$^{\rm -1}$}}

\newcommand{\msun}{\mbox{$M_{\odot}$}}
\newcommand{\logg}{\mbox{log~{\it g}}}

\newcommand{\RN}[1]{%
  \textup{\uppercase\expandafter{\romannumeral#1}}%
}
\DeclareMathOperator*{\argmin}{argmin}

\newcommand{\MnorE}{$0.62^{+0.12}_{-0.11}\times10^{11}$~\msun}
\newcommand{\rscaE}{$3.86^{+0.35}_{-0.38}$~kpc}
\newcommand{\slopE}{$0.91^{+0.04}_{-0.05}$}
\newcommand{\MvirE}{$1.81^{+0.06}_{-0.05}\times10^{11}$~\msun}
\newcommand{\rvirE}{$119.35^{+1.37}_{-1.21}$~kpc}
\newcommand{\concE}{$13.02^{+0.11}_{-0.10}$}
\newcommand{\rhoSE}{$0.447^{+0.004}_{-0.004}$~GeV\,cm$^{-3}$}
\newcommand{\JfacE}{$15.8^{+1.08}_{-0.93}\times10^{22}$~GeV$^{2}$\,cm$^{-5}$}
\newcommand{\chisE}{$2.97$}

\newcommand{\Mnorg}{$3.21^{+0.07}_{-0.07}\times10^{11}$~\msun}
\newcommand{\rscag}{$5.26^{+0.15}_{-0.11}$~kpc}
\newcommand{\slopg}{$0.0258^{+0.0416}_{-0.0192}$}
\newcommand{\Mvirg}{$6.94^{+0.12}_{-0.11}\times10^{11}$~\msun}
\newcommand{\rvirg}{$186.81^{+1.07}_{-1.04}$~kpc}
\newcommand{\concg}{$18.02^{+0.18}_{-0.18}$}
\newcommand{\rhoSg}{$0.405^{+0.001}_{-0.001}$~GeV\,cm$^{-3}$}
\newcommand{\Jfacg}{$33.1^{+1.38}_{-1.08}\times10^{22}$~GeV$^{2}$\,cm$^{-5}$}
\newcommand{\chisg}{$7.79$}

\title[The dark matter profile of the Milky Way]{The dark matter profile of the Milky Way inferred from its circular velocity curve}

\author[X. Ou et al.]{
Xiaowei Ou,$^{1}$\thanks{E-mail: xwou@mit.edu}
Anna-Christina~Eilers,$^{1}$\thanks{Pappalardo Fellow}
Lina Necib,$^{1,2}$
and Anna Frebel$^{1}$
\\
$^{1}$Physics Department and Kavli Institute for Astrophysics and Space Research,
Massachusetts Institute of Technology,\\
77 Massachusetts Ave, Cambridge MA 02139, USA\\
$^{2}$The NSF AI Institute for Artificial Intelligence and Fundamental Interactions,
Massachusetts Institute of Technology,\\
77 Massachusetts Ave, Cambridge MA 02139, USA
}

\date{Accepted XXX. Received YYY; in original form ZZZ}

\pubyear{2023}

\begin{document}
\label{firstpage}
\pagerange{\pageref{firstpage}--\pageref{lastpage}}
\maketitle

\begin{abstract}
In this paper, we construct the circular velocity curve of the Milky Way out to $\sim 30$\,kpc, providing an updated model of the dark matter density profile. We derive precise parallaxes for $120,309$ stars with a data-driven model, using \textit{APOGEE} DR17 spectra combined with \Gaia DR3, \textit{2MASS}, and \textit{WISE} photometry. At outer galactic radii up to $30$\,kpc, we find a significantly faster decline in the circular velocity curve compared to the inner parts. This decline is better fit with a cored Einasto profile with a slope parameter \slopE\ than a generalized Navarro-Frenk-White (NFW) profile. The virial mass of the best-fit dark matter halo profile is only \MvirE, significantly lower than what a generalized NFW profile delivers. We present a study of the potential systematics, affecting mainly large radii. 
Such a low mass for the Galaxy is driven by the functional forms tested, given that it probes beyond our measurements. It is found to be in tension with mass measurements from globular clusters, dwarf satellites, and streams. Our best-fit profile also lowers the expected dark matter annihilation signal flux from the galactic centre by more than an order of magnitude, compared to an NFW profile-fit. In future work, we will explore profiles with more flexible functional forms to more fully leverage the circular velocity curve and observationally constrain the properties of the Milky Way's dark matter halo.

\end{abstract}




\section{Introduction}

The rotation/circular velocity curve of a disc galaxy represents how fast an object would move at a given radial distance from the centre of the galaxy, assuming it is in a perfectly circular orbit. It measures the galaxy's mass as a function of the radial distance and has led to one of the strongest pieces of evidence for the unseen dark matter (DM) halos surrounding almost all extragalactic galaxies we observe \citep{rubin80}. 

The circular velocity curve has long been used within our galaxy to constrain the Milky Way mass and mass distribution. Specifically, since the masses of galaxies like our Milky Way are primarily made up of DM \citep{faber79}, the dynamics of luminous components are dominated by the DM potential, providing an indirect probe to the DM. The underlying DM density profile can thus be inferred from the circular velocity curve. Such information is crucial for both direct and indirect detection of DM: first, the local DM density at the solar position is directly proportional to the expected rate of DM direct detection events \citep{Goodman85,Drukier86,1996PhR...267..195J}. Second, the DM profile in the inner part of the Galaxy is important in estimating the integrated DM density at the galactic centre, which is key in DM indirect detection searches \citep{abdallah16,ackermann17,abeysekara18,acharyya21}.

The circular velocity curve has been measured with various tracers, ranging from molecular clouds to masers to bright stellar tracers: At galactocentric radii within the solar radius, the tangent-point method derives the rotation curve from measuring the radio emission from H\,\textsc{i} and CO lines of the interstellar medium, assuming the gas moves in purely circular orbit \citep{gunn79,fich89,levine08,sofue09}. At galactocentric radii outside of the solar radius, tracers with relatively easily measurable distances, proper motions, and(or) line-of-sight velocities have been used to constrain the rotation curve. Namely, the stellar standard candles of classical cepheids \citep{pont97}, red clump giants \citep{bovy12,huang16}, RR Lyrae stars \citep{ablimit17,wegg19}, and blue horizontal branch stars \citep{xue09,kafle12} with radial velocity measurements are viable tracers but are often rare or not bright enough to be observable at large distances. Other non-stellar tracers such as the thickness of the H\,\textsc{i} layer \citep{merrifield92}, spectrophotometric distances of H\,\textsc{ii} regions combined with radial velocities of associated molecular clouds \citep{fich89,brand93}, planetary nebulae \citep{schneider83}, and masers in high-mass star-forming regions \citep{reid14} are also either rare or indirectly connected with the rotation curve through simplifying modelling and assumptions.

The advent of large astrometric surveys like \Gaia\ has greatly expanded the range of high precision parallax and proper motion measurements for stars \citep{gaia16,gaia18,gaia21}. Combined with line-of-sight velocity measurements from large spectroscopic surveys such as \textit{APOGEE} \citep{majewski17}, the circular velocity curve of the Galaxy can be mapped out to large galactocentric distances. One limiting factor, however, is the precision of the parallax measurements at heliocentric distances greater than $\sim 5$\,kpc. 

Efforts have been made to improve the precision of astrometric parallaxes/distances measured by \Gaia\ with the goal of better constraining the circular velocity curve to large heliocentric distances ($>10$\,kpc). Previous studies by \citet{hogg19} have demonstrated that using a simple data-driven linear model trained on photometric and spectroscopic features of luminous RGB stars can help greatly improve the precision of the astrometric parallaxes. In \citet{eilers19}, these new parallaxes yielded a much tighter constraint on the circular velocity curve of the Milky Way out to galactocentric distances of $\sim 25$\,kpc. \citet{wang22} applied a statistical deconvolution of the parallax errors based on Lucy's inversion method (LIM) to the \Gaia\ third data release (DR3) sources and obtained the circular velocity curve of the Milky Way out to $\sim 30$\,kpc. \citet{zhou22} used a supervised machine learning algorithm trained on \Gaia\ astrometric distances to predict distances to luminous bright RGBs based on photometric and spectroscopic features.

In this study, we present an updated circular velocity curve out to galactocentric radius $\sim 30$\,kpc using a similar procedure as performed in \citet{eilers19}. With the new data from \Gaia\ DR3 and \textit{APOGEE} DR17, we are able to measure the curve to a further distance with higher precision. We study the implication of the new improved circular velocity curve on the Milky Way DM density profile. Specifically, we briefly describe the data set used for this study in Section~\ref{sec:data}. We lay out the process of deriving precise parallaxes with a data-driven model in Section~\ref{sec:sp_plx}, using \textit{APOGEE} DR17 spectra combined with photometry measurements from \Gaia, \textit{2MASS}, and \textit{WISE}. Section~\ref{sec:vcirc_curve} covers the assumption and model used to measure the circular velocity curve of the Milky Way out to $\sim 30$\,kpc using the spectrophotometric parallaxes. The DM profile analysis procedure is described in Section~\ref{sec:dm_analysis}, and the results are shown in Section~\ref{sec:result}. We discuss the implications of our results in Section~\ref{sec:discussion} and conclude with Section~\ref{sec:conclusion}.

\section{Data}
\label{sec:data}

We use luminous red giant branch (RGB) stars as tracers for measuring the circular velocity curve in this study. 
RGB stars are an ideal tracer of the galactic disc due to their high luminosities and, thus, large observable volume. It is also principally possible to predict the luminosity of an RGB star given its spectroscopy (for stellar parameters) and photometry (for extinction correction) measurements. Their luminosities are simple functions of their composition, surface gravity, temperature, and age, which can be derived by spectroscopic observation. Their location on a colour-magnitude diagram is photometrically near-orthogonal to reddening vectors by dust and, thus, can be relatively easily de-reddened. 

In this study, we utilize a data-driven model to predict RGB star parallaxes using photometric and spectroscopic measurements, which we describe in more detail in Section~\ref{sec:sp_plx}. We refer to the predicted parallaxes as the spectrophotometric parallaxes, in contrast with the astrometric parallaxes measured by \Gaia. Based on the physical expectation listed above, we allow the data-driven model to learn patterns in a given data set and discover the relationships between spectral features in the spectra of the stars, photometry (including colours), and parallax (or distance). For spectroscopic observation, we take spectra from \textit{APOGEE} DR17 \citep{majewski17}. For photometry, we combine measurements from \Gaia\ DR3 \citep{gaia21}, \textit{WISE} \citep{wright10}, and \textit{2MASS} \citep{skrutskie06}. Specifically, we include photometric magnitudes in $G$, $G_{BP}$, $G_{RP}$, $W_1$, $W_2$, $J$, $H$, and $K$ bands for this study, following \citet{hogg19}.

We select RGB stars from \textit{APOGEE} DR17 by requiring surface gravity (\logg) between $0.0$ and $2.2$. This cut selects all stars more luminous than the red clump stars. While the range of \logg\ spans over two dex, the derived circular velocity curve and the DM profile result do not show systematic differences as a function of \logg.

We crossmatch the selected RGB stars from \textit{APOGEE} with \Gaia\ DR3. \textit{WISE} and \textit{2MASS} photometries are pre-matched with \Gaia\ and \textit{APOGEE}, respectively, in the data releases and taken as is. We additionally apply two quality cuts on colours, as described in \citet{hogg19}, to remove stars with outlying photometry. These cuts remove $2.5\%$ of the \textit{APOGEE} sample. 
The final parent sample suitable for spectrophotometric parallax calculation contains 120,309 stars, nearly tripled compared to the previous study with \textit{APOGEE} DR14 with 44,784 stars \citep{eilers19}.

\section{Spectrophotometric Parallaxes}
\label{sec:sp_plx}

We follow the same procedure from \citet{hogg19} to derive the spectrophotometric parallaxes. We briefly review the methodology here, but readers are encouraged to go through the original paper for more details.

The data-driven model fundamentally assumes that the parallaxes of all RGB stars selected can be completely described by photometric and spectroscopic information using a linear model. Specifically, the model assumes that the logarithm of the true parallax can be expressed as a linear combination of the components of a $D$-dimensional feature vector $\vec{x}_n$ and a $D$-dimensional coefficient vector $\vec{\theta}$. The model is thus expressed by
\begin{equation}
    \varpi_{n}^{(a)} = \exp{(\vec{\theta} \cdot \vec{x}_n)} + \text{noise}.
\end{equation}
$\varpi_{n}^{a}$ is the astrometric parallax measurement from \Gaia\ of star $n$. The feature vector $\vec{x}_n$ contains the photometric magnitudes and spectroscopic normalized\footnote{Spectra are normalized using spectral analysis tools developed by A.~Ji: \url{https://github.com/alexji/alexmods}.} fluxes and thus consists of eight photometric features ($G$, $G_{BP}$, $G_{RP}$, $J$, $H$, $K$, $W_1$, and $W_2$) and $7,451$ spectroscopic features (after removing CCD chip gaps and flagged bad pixels from the 8575-pixel \textit{APOGEE} spectra). This results in $D=7460$, where we also added one additional constant term. We apply a constant offset of $\Delta \varpi_{n}^{(a)} = 17\,\mu$as to all astrometric parallaxes to account for the reported median parallax bias in Gaia DR3 \citep{lindegren21}.

In order to optimize the coefficient vector, we adopt the log-likelihood function
\begin{equation}
    \log{\mathcal{L}} = - \frac{1}{2} \chi^2(\vec{\theta}) = - \sum^{N}_{n=1} \frac{[\varpi_{n}^{(a)} - \exp{(\vec{\theta} \cdot \vec{x}_n)}]^2}{2\sigma_{n}^{(a)2}},
\end{equation}
where $\sigma_{n}^{(a)}$ is the uncertainty on \Gaia\ parallax. As described in \citet{hogg19}, we do not expect all spectral pixels in the normalized spectra to contain information about the physical properties of the star according to the sparsity assumption. Only a small subset of the full set of \textit{APOGEE} spectral pixels will provide information for the prediction of parallax. We thus apply an additional term (regularization term) to the likelihood function and optimize the regularized objective function (Equation~\ref{eq:reg_func}) to account for the sparsity assumption. This function essentially allows the model to under-fit in regions of the spectra that do not contain spectral features relevant to the luminosity and distance of the star. 
The regularized objective function,
\begin{equation}
    \hat{\vec{\theta}} \longleftarrow \argmin_{\vec{\theta}} [\frac{1}{2} \chi^2 (\vec{\theta}) + \lambda ||P\cdot\vec{\theta}||^1_1],
    \label{eq:reg_func}
\end{equation}
introduces $\lambda$, the regularization parameter, and $P$, a projection operator that selects only features corresponding to the \textit{APOGEE} spectral pixels, i.e. the regularization only applies to the spectral pixels. 

\begin{figure*}
    \centering
    \includegraphics[width=\textwidth]{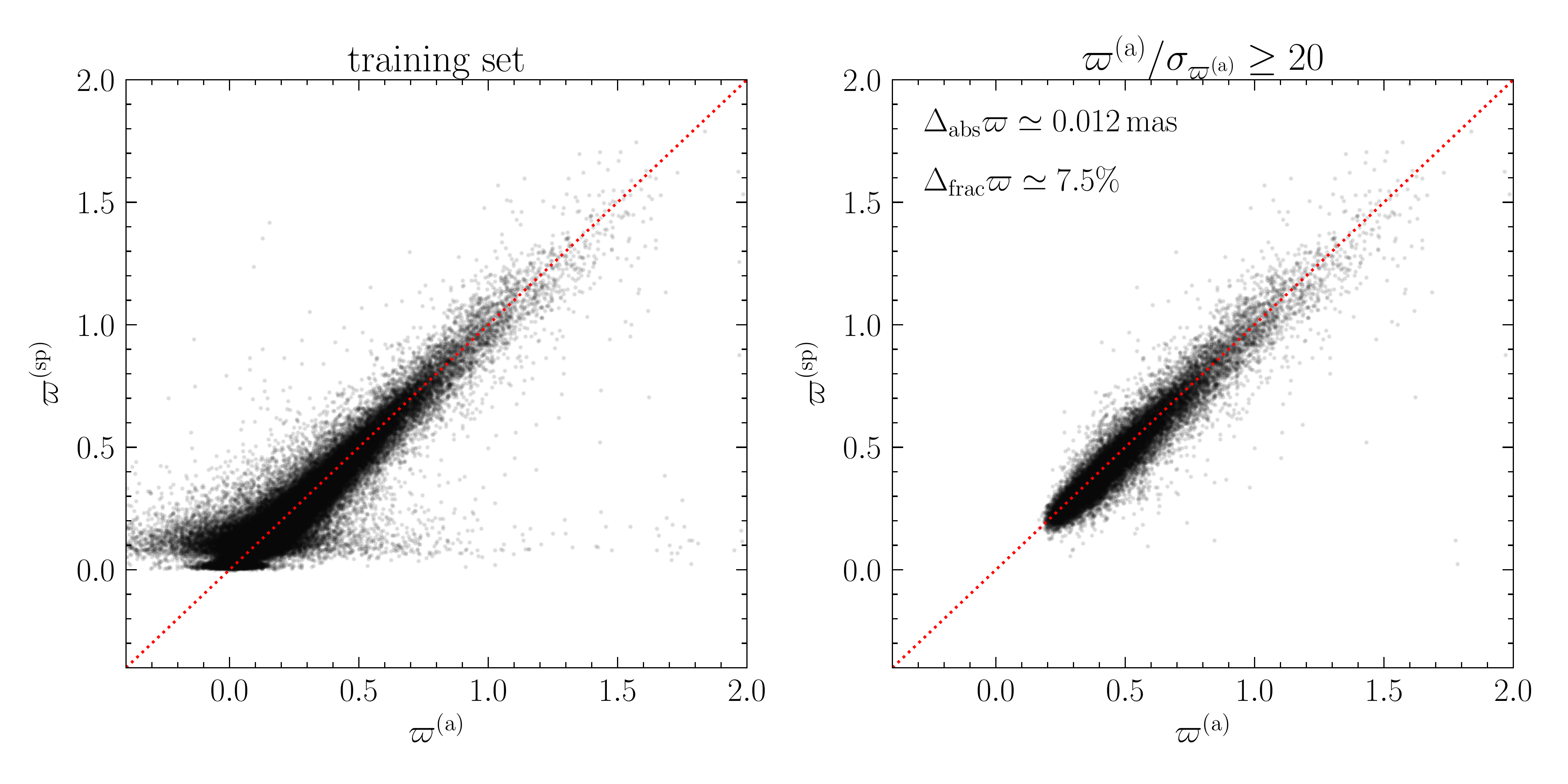}
    \caption{Comparison between the Gaia astrometric parallax (horizontal axis) and our spectrophotometric parallaxes (vertical axis). The left panel shows the full training sample, whereas the right panel includes only the stars with high signal-to-noise measurements from Gaia (\texttt{parallax\_over\_error} $> 20$). The median fractional difference between the two parallax measurements for the high signal-to-noise subset is $\sim 7.5\%$, and the median absolute difference is $\sim 0.012$~mas.}
    \label{fig:plx_compare}
\end{figure*}

We split the parent sample randomly into two sub-samples (A and B) and use sample A(B) as the training set for sample B(A). Using the optimized coefficient vector $\hat{\vec{\theta}}$ from the training set, we infer the spectrophotometric parallax estimate ($\varpi_{m}^{(sp)}$) for each star in the validation set and compare $\varpi_{m}^{(sp)}$ with the \Gaia\ astrometric parallaxes ($\varpi^{(\rm{a})}$). The value of $\lambda$ is set to 140 via cross-validation, i.e. we vary $\lambda$ from $10$ to $240$ to find where the fractional differences between the $\varpi^{(\rm{a})}$ and $\varpi^{(\rm{sp})}$ based on the training set is at a minimum. With the final adopted $\lambda=140$, the median absolute difference between the astrometric and spectrophotometric parallaxes ($\Delta_{\rm{abs}} \varpi = \rm{Med}(|\varpi^{(\rm{sp})}-\varpi^{(\rm{a})}|)$) is $\sim 0.012$~mas for stars with \texttt{parallax\_over\_error} larger than 20 from \Gaia, whereas the fractional difference ($\Delta_{\rm{frac}} \varpi= \rm{Med}(|\varpi^{(\rm{a})}-\varpi^{(\rm{a})}|/\varpi^{(\rm{a})})$) is approximately 7.5\%. Figure~\ref{fig:plx_compare} shows the agreement between the two, both for the full sample and for the subset of stars with high signal-to-noise astrometric parallax measurements from \Gaia.

We obtain a median relative uncertainty in spectrophotometric parallax $\sim 8 \%$. Compared to $\sim 14 \%$ for \Gaia\ parallax of the full sample used in this study, our results show $\sim 40\%$ improvement.
At a heliocentric distance greater than 3(18)\,kpc, the spectrophotometric parallaxes are approximately 2.5(10) times as precise as \Gaia\ parallaxes, which allows us to map the circular velocity curve out to a radial galactic distance of 30\,kpc.

\section{Circular velocity curve}
\label{sec:vcirc_curve}

We use the estimated spectrophotometric parallaxes, discussed in Section~\ref{sec:sp_plx}, to derive the circular velocity curve.
To this end, we apply the coordinate transformation to galactocentric coordinates assuming a distance from the Sun to the galactic centre of 8.178\,kpc \citep{gravity19}, a height of the Sun above the galactic plane of 0.0208\,kpc \citep{bennett19}, and solar galactocentric velocities of $(v_x, v_y, v_z)\sim(5.1,247.3,7.8)$\,\kmsec \citep{reid04,schonrich10,gravity19}.
Uncertainties in proper motion, spectrophotometric parallax, and radial velocities are propagated to the final galactocentric positions and velocities. Correlations between proper motions, as reported by \Gaia, are included.

\begin{figure*}
    \centering
    \includegraphics[width=\textwidth]{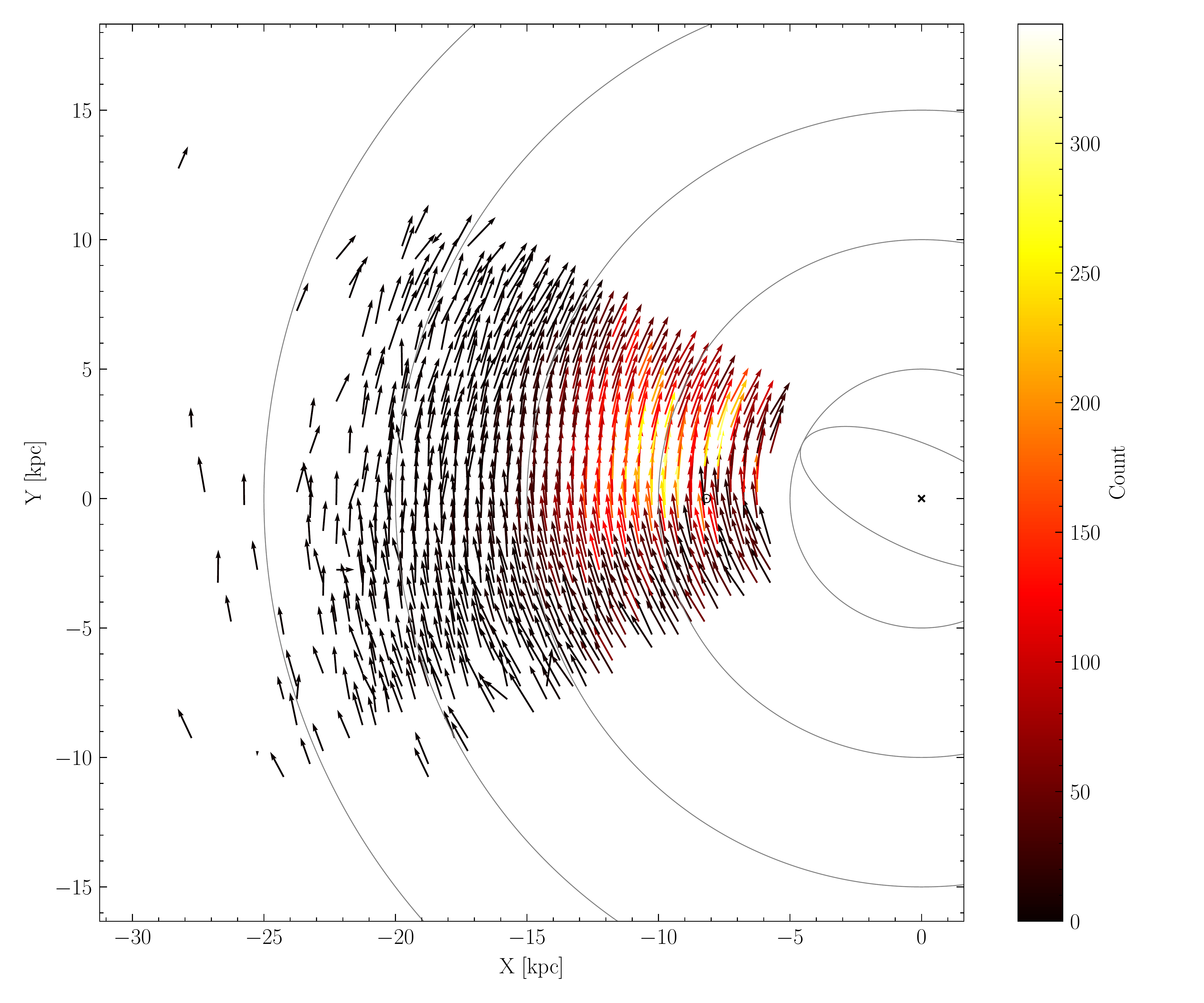}
    \caption{Galactocentric XY-plane map of the $33,335$ stars used for calculating circular velocities, plotted in 0.5\,kpc bins. The vectors represent the mean velocity of stars within each bin, colour-coded by the number of stars in each bin.}
    \label{fig:XYmap}
\end{figure*}

We select disc stars with the same cuts used in \citet{eilers19}. Namely, we require that the $\alpha$-element abundances measured by \textit{APOGEE} $[\alpha/\rm{Fe}] < 0.12$ to avoid large asymmetric drift corrections. To remove contamination from the halo and account for a possible flaring in the outer disc, we select stars with velocity perpendicular to the galactic plane $|v_z| < 100$\,\kmsec\ in velocity space and height above the galactic plane $|z| < 1$\,kpc or within 6\,\degree\ from the galactic plane $|z|/R < \tan{\pi/30}$ in position space. \footnote{We follow the standard notations $(r,\phi,\theta)$ and $(R,\phi,z)$ for cylindrical and spherical coordinates, respectively, in the galactocentric frame, unless otherwise noted.}
We also limit our sample to within a wedge of 60° from the galactic centre toward the direction of the Sun and remove stars potentially affected by the non-axisymmetric potential near the galactic bar at $R < 6$\,kpc. The final sample size for calculating circular velocity is $33,335$, $\sim 50\%$ more than the previous study with 23,129 stars \citep{eilers19}. Figure~\ref{fig:XYmap} shows the vector map of the final disc sample used for calculating the circular velocity curve. The sample populates well out to $R \sim 25$\,kpc, and sparsely out to $R \sim 30$\,kpc (20 stars in total at $R>25$\,kpc).

Assuming the galactic potential outside of $R\sim 6$\,kpc is axisymmetric, we use Jeans' equation to measure the circular velocity curve with this sample via
\begin{equation}
    \frac{\partial \nu \langle v^2_R\rangle}{\partial R} + \frac{\partial\nu\langle v_R v_z\rangle}{\partial z} + \nu\left(\frac{\langle v^2_R\rangle - \langle v^2_{\varphi}\rangle}{R} + \frac{\partial\Phi}{\partial R} \right) = 0,  
    \label{eq:jeans}
\end{equation}
where $\nu$ is the density distribution of the tracer population; we approximate the radial profile of the tracer population by an exponential function with a scale length of 3\,kpc due to a lack of knowledge of the selection function. The choice of functional form and associated parameter(s) can induce systematic uncertainties up to $\sim 2 \%$ in the final circular velocity curve measurements \citep{eilers19}.

The circular velocity is then calculated using 
\begin{equation}
    v^2_{\rm c}(R) = \langle v^2_{\varphi}\rangle - \langle v^2_R\rangle\left(1 + \frac{\partial \ln \nu}{\partial \ln R} + \frac{\partial \ln \langle v^2_R\rangle}{\partial \ln R}\right), 
    \label{eq:vcirc2}
\end{equation}
where the second term in Equation~\ref{eq:jeans} is omitted for it is $\sim 2-3$ orders of magnitude smaller than other terms in the equation and introduces systematic uncertainties only at the $\sim 1 \%$ level \citep{eilers19}.

\begin{figure}
    \centering
    \includegraphics[width=0.45\textwidth]{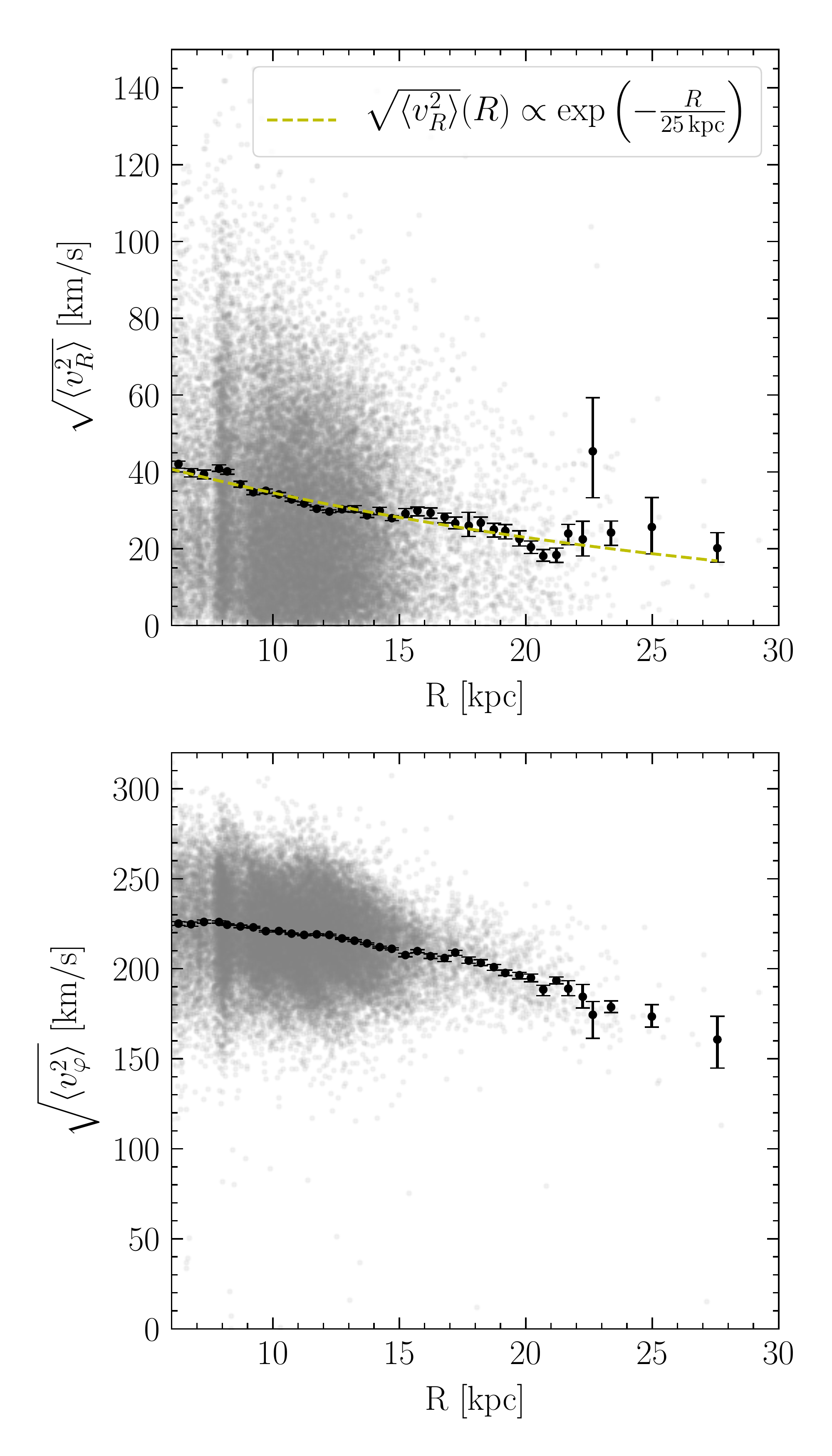} 
    \caption{Radial profiles of the components of Equation~\ref{eq:jeans} ($\sqrt{\langle v^2_{R}\rangle}$ (top) and $\sqrt{\langle v^2_{\varphi}\rangle}$ (bottom)). Grey dots in the background represent individual stars. Black dots are the ensemble averages of the stars in the same $R$ bins used in calculating the circular velocities, with the uncertainties estimated via bootstrapping with 100 samples. The fitted dependency of $\sqrt{\langle v^2_{R}\rangle}$ is shown with the yellow dashed curve in the top panel.}
    \label{fig:v_disp}
\end{figure}

The calculation is carried out in the same way described in \citet{eilers19}, to which readers may refer for more details. 
We reiterate a few key ingredients here. The radial density profile is modeled by an exponential function with a fixed scale length of 3\,kpc, consistent with recent studies \citep{bland-hawthorn16}. 
The radial velocity tensor ($\langle v^2_R\rangle$) profile is also modeled by an exponential function but with an estimated scale length of $\sim25$\,kpc based on the data, as shown in Figure~\ref{fig:v_disp}.
We report the final radial velocities measurements in Table~\ref{tab:vcirc}. We note that $R$ is not exactly evenly spaced but rather calculated from the weighted mean of stars in each bin. We adjust the bin size at larger $R$ to improve the statistics within each bin, i.e. each bin contains at least five stars. 

\begin{figure*}
    \centering
    \includegraphics[width=\textwidth]{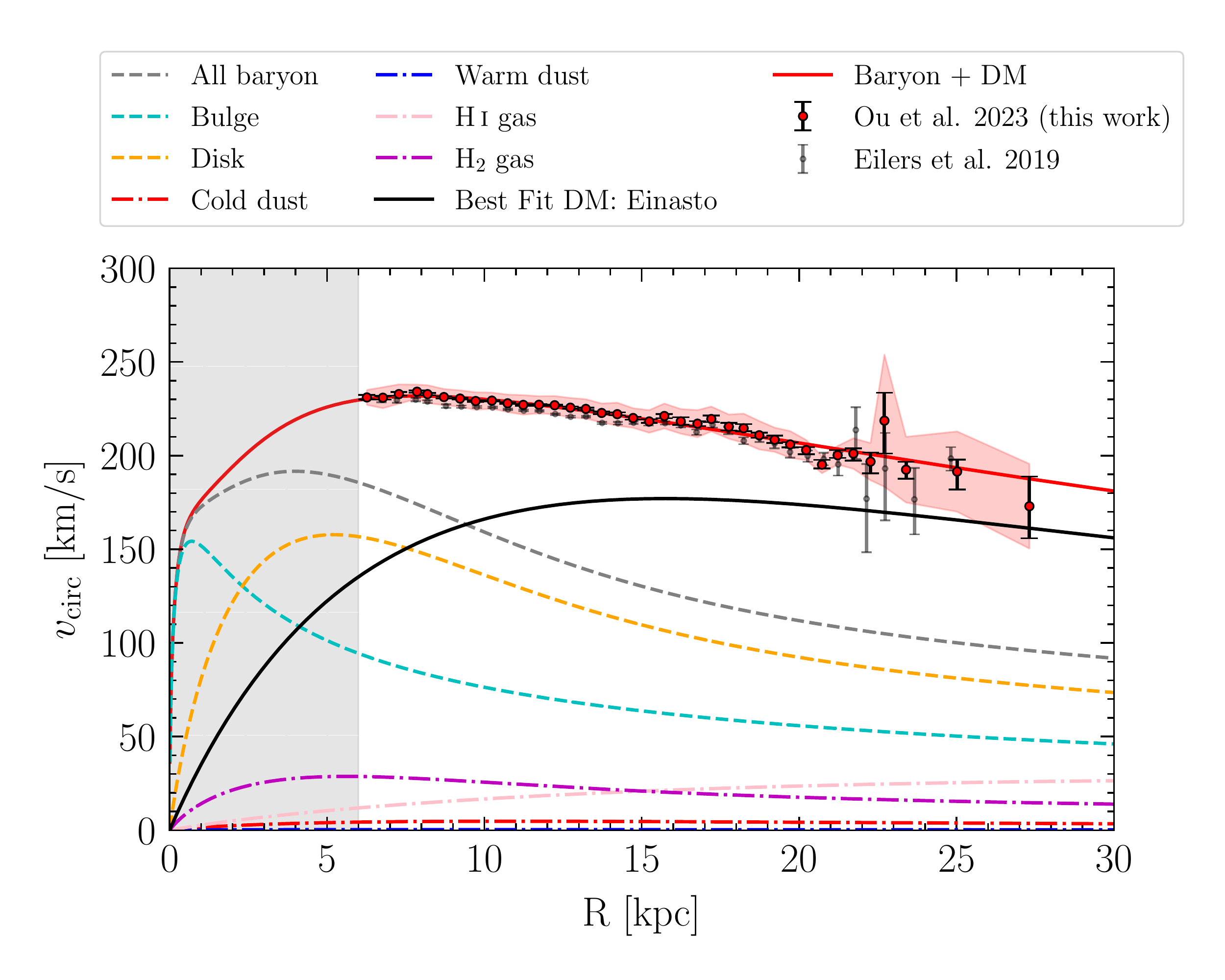} 
    \caption{Comparison between the circular velocity curve measured from \citet{eilers19} (black) and this work (red). The best-fit Einasto DM profile, with the baryonic model from \citet{desalas19}, is also shown here. The grey shaded region represents the bulge region, which we do not model due to the non-axisymmetric potential near the galactic bar.
    The red shaded region represents the total uncertainty estimate from the dominating systematic sources, as shown in Figure~\ref{fig:sys_err}.
    }
    \label{fig:vcirc_compare}
\end{figure*}

Figure~\ref{fig:vcirc_compare} shows our final circular velocity curve. We observe a shallow and steady decline in the curve, from $234.14$\,\kmsec\ at $R=7.86$\,kpc to $172.98$\,\kmsec\ at $R=27.31$\,kpc. The curve is smooth and declines slowly between $R=7$-$17$\,kpc with a difference in velocities of approximately 20\,\kmsec. The velocity drops faster between $R=17$-$27$\,kpc with a difference of 40\,\kmsec.

\begin{figure*}
    \centering
    \includegraphics[width=0.8\textwidth]{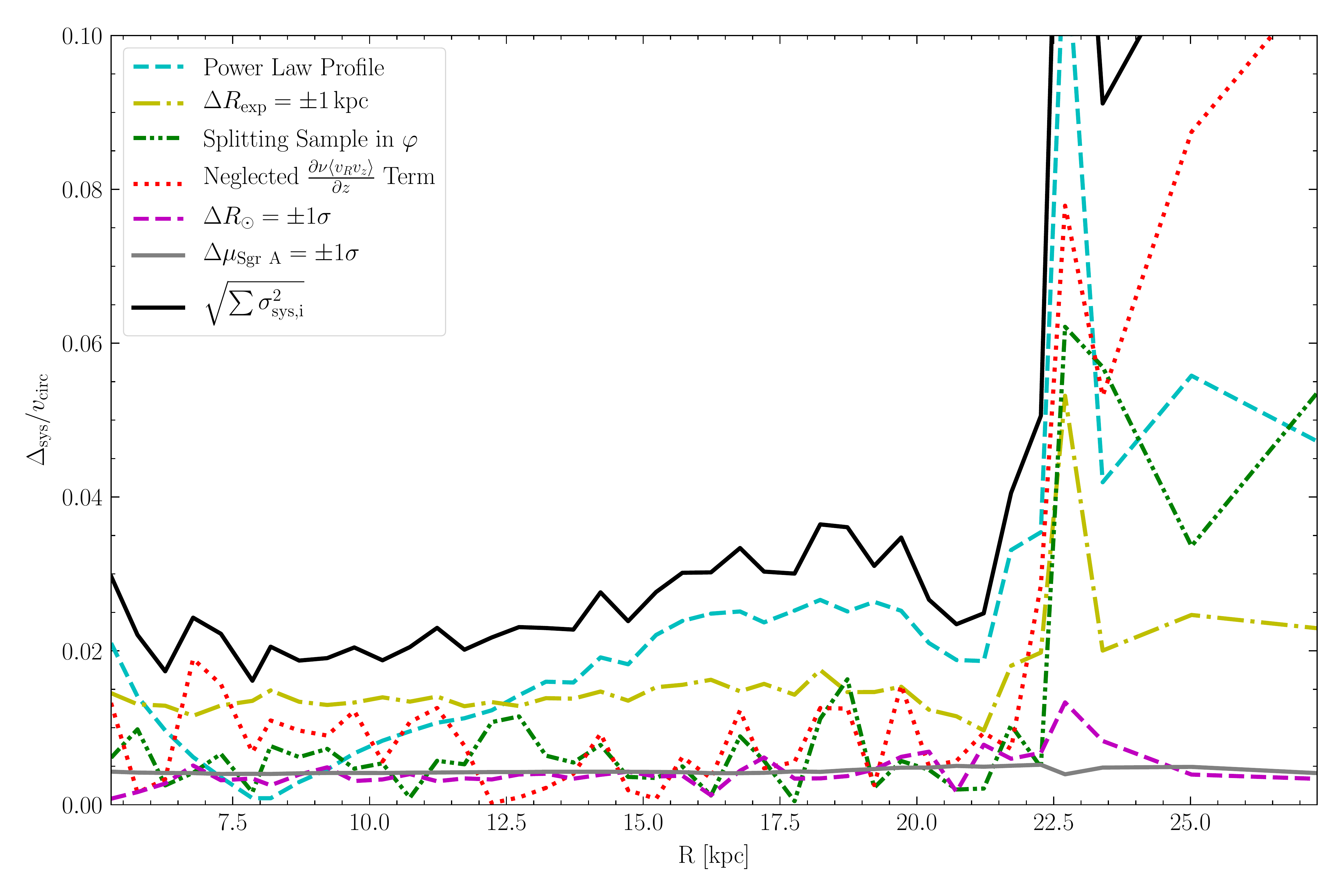}
    \caption{
    Summary of systematic uncertainties in the circular velocity curve. The systematic uncertainties from assuming a power law with an index of $-2.7$ instead of an exponential function for the radial density profile of the tracer population (cyan dashed line), from varying the scale length in the exponential radial density profile (yellow dashed-dotted line), and from splitting the sample into two distinct wedges (green dashed-double-dotted line) are moderate between $\sim1\%$ to $\sim3\%$. Similarly, we see very moderate uncertainties from the assumed solar distance from the galactic centre (magenta dashed line) and the proper motion of Sgr A$^{*}$ (grey solid line) at $<1\%$ level. The uncertainty from the neglected term in Equation~\ref{eq:jeans} (red dotted line) is very moderate ($<1\%$) up to $R=20$\,kpc but increases and dominates at larger radii. 
    The total systematic uncertainty (black solid line) ranges between $1\%$ to $5\%$ up to $R = 22$\,kpc, increasing to $15\%$ due to the large asymmetric drift correction as described in Section~\ref{sec:vcirc_curve}.
    }
    \label{fig:sys_err}
\end{figure*}

In Figure~\ref{fig:sys_err}, we examine the systematic uncertainties that originated from (i) the assumed functional form for the radial density profile, (ii) the uncertainties in the exponential scale length of the radial density profile, (iii) splitting the sample into two distinct wedges, and (iv) the neglected asymmetric correction term in Equation~\ref{eq:jeans}.
The individual systematic uncertainties range between $1\%$ to $5\%$ up to $R = 22$\,kpc, with a total systematic uncertainty between $2-4\%$.
At $R > 22$\,kpc, the total systematic uncertainty is dominated by the neglected asymmetric drift correction term $\left(\frac{\partial\nu\langle v_R v_z\rangle}{\partial z}\right)$, reaching over $15\%$. 
Moreover, the outermost data points suffer from the lack of stars at that distance, preventing an accurate estimate of the neglected asymmetric drift correction term. 
It is, thus, difficult to properly apply the asymmetric drift correction at this galactic radius and fair to assume larger uncertainties at that distance. 

Comparing our result and those from \citet{eilers19} in Figure~\ref{fig:vcirc_compare}, we see good agreements at $R < 15$\,kpc, with a systematic offset of $\sim 5$\,\kmsec, which is of the same order as the systematic uncertainties in, i.e.\ different choices of solar position and velocity with respect to the galactic centre, tracer population scale length, and functional form of the tracer population (see Figure~\ref{fig:sys_err}). 
Additional systematic uncertainties on the $<1\%$ level \citet{eilers19} could arise from different values for the assumed solar distance from the galactic centre and the solar vertical height above the galactic plane, where \citet{eilers19} adopted values from \citet{gravity18} and \citet{juric08}, respectively, while in this work we use the more recent measurements from \citet{gravity19} and \citet{bennett19}. Our solar velocities with respect to the galactic centre are calculated similarly using proper motion measurements of Sgr A$^{*}$ from \citet{reid04} and the Solar motion along the line-of-sight to Sgr A$^{*}$ from \citet{schonrich10}. Nevertheless, with updated Solar distance and correction on the Solar motion along the line-of-sight to Sgr A$^{*}$, we adopt different solar velocities from \citet{eilers19}.

At large $R >15$\,kpc, our result agrees with \citet{eilers19} within 1-$\sigma$, but is much smoother and extends to larger $R$ with smaller uncertainties. Most importantly, we see more clearly a steady and fast decline in the curve at outer galactic radii ($R >20$\,kpc). Note that the decline is unlikely to be a result of contaminants from the stellar halo (see Appendix~\ref{sec:contaminant}).

\begin{table}
\caption{Measurements of the Circular Velocity of the Milky Way.}
\label{tab:vcirc}
\begin{tabular}{rrccr}
\hline
$R$ & 
$v_c$ & 
$\sigma^{+}_{v_c}$ & 
$\sigma^{-}_{v_c}$ & 
$N_{\rm{star}}$ \\
\hline
[kpc] & 
[\kmsec)] & 
[\kmsec)] & 
[\kmsec)] & 
  \\
\hline

6.27 & 231.07 & 1.28 & 1.00 & 764 \\
6.78 & 230.93 & 0.97 & 0.97 & 676 \\
7.28 & 232.87 & 1.13 & 0.79 & 812 \\
7.86 & 234.14 & 0.63 & 0.62 & 1631 \\
8.19 & 232.83 & 0.56 & 0.55 & 2270 \\
8.71 & 231.27 & 0.54 & 0.66 & 1445 \\
9.23 & 230.47 & 0.44 & 0.46 & 2179 \\
9.72 & 229.16 & 0.54 & 0.43 & 2505 \\
10.24 & 229.37 & 0.25 & 0.50 & 2560 \\
10.74 & 227.95 & 0.53 & 0.34 & 2528 \\
11.23 & 227.09 & 0.44 & 0.46 & 2692 \\
11.73 & 227.15 & 0.35 & 0.45 & 2419 \\
12.23 & 226.90 & 0.39 & 0.41 & 2285 \\
12.73 & 225.61 & 0.64 & 0.53 & 1994 \\
13.22 & 224.95 & 0.66 & 0.69 & 1665 \\
13.72 & 222.79 & 0.68 & 0.53 & 1308 \\
14.22 & 222.13 & 1.04 & 0.64 & 938 \\
14.73 & 220.08 & 0.65 & 0.89 & 641 \\
15.24 & 218.25 & 1.14 & 0.79 & 449 \\
15.72 & 221.16 & 1.06 & 1.24 & 322 \\
16.24 & 218.30 & 2.17 & 1.88 & 243 \\
16.77 & 217.07 & 1.39 & 1.35 & 164 \\
17.21 & 219.56 & 1.80 & 1.69 & 150 \\
17.77 & 215.49 & 1.98 & 2.08 & 114 \\
18.23 & 214.62 & 2.24 & 1.59 & 102 \\
18.73 & 210.89 & 1.42 & 1.32 & 94 \\
19.22 & 208.48 & 2.20 & 1.65 & 71 \\
19.71 & 205.97 & 1.20 & 1.55 & 70 \\
20.22 & 202.97 & 1.55 & 2.25 & 65 \\
20.72 & 195.16 & 2.60 & 1.97 & 46 \\
21.22 & 200.20 & 2.84 & 1.45 & 38 \\
21.72 & 201.11 & 2.72 & 3.81 & 30 \\
22.27 & 196.79 & 4.82 & 6.25 & 14 \\
22.71 & 218.65 & 14.93 & 17.54 & 11 \\
23.40 & 192.49 & 4.25 & 4.8 & 22 \\
25.02 & 191.48 & 6.41 & 9.61 & 11 \\
27.31 & 172.98 & 15.82 & 17.07 & 7 \\
\hline
\end{tabular}
\end{table}

\section{Dark Matter profile Analysis}
\label{sec:dm_analysis}

We model the circular velocity curve obtained above as a result of gravitational potential composed of baryonic and DM components. For simplicity of the analysis, we assume the baryonic potential is fixed and derive the DM potential necessary to reproduce the measured circular velocity curve. Future work will include a discussion of different baryonic potentials.

The baryonic potential for the rest of the study is chosen and fixed to replicate the model B2 described in \citet{desalas19}. The model (and the relevant parameters) is primarily based on studies from \citet{misiriotis06}, whereas the bulge model is proposed in \citet{desalas19}. This model is designed to address issues with the overestimated mass of the baryons towards the outer galactic radii from model B1 in the same study (also the one used originally in \citealt{eilers19}), which was taken from \citet[model I]{pouliasis17}. Thus, by design, the baryonic potential adopted in this study is overall less massive than what was used in \citet{eilers19}. 
We describe the model in more detail below.

The model comprises six axisymmetric components: two stellar components for the disc and the bulge, two dust components (cold and warm), and the molecular H$_2$ and atomic H\,\textsc{i} gas. All except for the bulge are modelled as double exponential profiles expressed as 
\begin{equation}
    \rho (R,z) = \frac{M_0}{4 \pi z_d R_d^2} \exp{\left(-\frac{R}{R_d}-\frac{|z|}{z_d}\right)},
    \label{eq:dens_doubexp}
\end{equation}
where $M_0$ is the mass normalization, and $R_d$ and $z_d$ are the scale length and height, respectively. \citet{mancera22} has shown that gas disc flaring has minimal effect on circular velocity curve analysis in non-gas-dominated galaxies. We thus do not model potential flaring in the Milky Way gas disc.

The bulge is modelled with a Hernquist potential 
\begin{equation}
    \rho (r) = -\frac{G M_0}{r_b + r},
    \label{eq:dens_hernquist}
\end{equation}
where $M_0$ is the mass normalization and $r_d$ is the scale radius. The relevant parameters for the double exponential profiles are taken from \citet{misiriotis06}, whereas the bulge profile parameters are taken from \citet{desalas19}. Values of the parameters are summarized in Table~\ref{tab:bary_models}.

\begin{table*}
\caption{Input parameters for the baryonic model.}
\label{tab:bary_models}
\begin{threeparttable}[t]
\centering
\begin{tabular}{lccc}
\hline
  & 
Normalization Mass ($M_0$) & 
Scale Length ($R_d$, $r_b$) & 
Scale Height ($z_d$) \\
\hline
 & 
[\msun] & 
[kpc] & 
[kpc] \\
\hline

\textbf{Disk}\tnote{a} & $3.65\times10^{10}$ & 2.35 & 0.14 \\
\textbf{Warm Dust}\tnote{a} & $2.20\times10^{5}$ & 3.30 & 0.09 \\
\textbf{Cold Dust}\tnote{a} & $7.00\times10^{7}$ & 5.00 & 0.10 \\
\textbf{H\,I Gas}\tnote{a} & $8.20\times10^{9}$ & 18.24 & 0.52 \\
\textbf{H$_2$ Gas}\tnote{a} & $1.30\times10^{9}$ & 2.57 & 0.08 \\
\textbf{Bulge}\tnote{b} & $1.55\times10^{10}$ & 0.70 & -- \\

\hline
\end{tabular}
\begin{tablenotes}
     \item[a] Double exponential profiles \citep{misiriotis06}.
     \item[b] Hernquist profile \citep{desalas19}.
\end{tablenotes}

\end{threeparttable}
\end{table*}

We apply the Markov Chain Monte Carlo affine invariant sampler \textsc{emcee} \citep{foremanmackey13} to fit the DM halo models. We fit two models, generalized-NFW (gNFW) and Einasto profiles, separately to test how well each can recover the declining behaviour of the circular velocity at outer galactic radii. The gNFW profile is a generalization to the well-known NFW profile, which is a common approximation to DM density profiles found in cosmological simulations \citep{navarro97}. Unlike the standard cuspy NFW profile, which diverges towards smaller $r$, the gNFW profile adds a free parameter that modulates the inner and outer asymptotic power law slope of the standard NFW profile, allowing it to be completely cored with a power law slope down to $-3$ in density at radii larger than the scale radius. The Einasto profile is also widely used to describe the density profile of galaxies \citep{einasto65,retana-montenegro12}. The functional form of an Einasto profile also allows for a cuspy or cored profile. Unlike the asymptotic power law behaviour for an NFW/gNFW profile, the Einasto profile has an exponential decrease in density outside the centre region. In light of the steady decrease we observe in the measured circular velocity curve, we test which model provides a better fit to the new data presented in this work. 

For the gNFW profile, we calculate the circular velocity curve based on the density profile of the form
\begin{equation}
    \rho_{\rm{gNFW}} (r) = \frac{M_0}{4 \pi r_s^3} \frac{1}{(r/r_s)^\beta(1+r/r_s)^{3-\beta}},
    \label{eq:dens_gNFW}
\end{equation}
where $M_0$ is the mass normalization, $r_s$ is the scale radius, and $\beta$ is the characteristic power for the inner part of the potential. When $\beta=1$, we recover the standard NFW profile. The Einasto profile is defined as
\begin{equation}
    \rho_{\rm{Ein}} (r) = \frac{M_0}{4 \pi r_s^3} \exp{\left(-(r/r_s)^\alpha\right)},
    \label{eq:dens_einasto}
\end{equation}
where $M_0$ and $r_s$ are defined similarly as in Equation~\ref{eq:dens_gNFW}, and $\alpha$ determines how fast the density distribution falls with galactic radius.

The model circular velocity at any given $R$ is then computed from the total enclosed mass, calculated by integrating the density profile. 

\section{Results}
\label{sec:result}

\begin{figure*}
    \centering
    \includegraphics[width=0.45\textwidth]{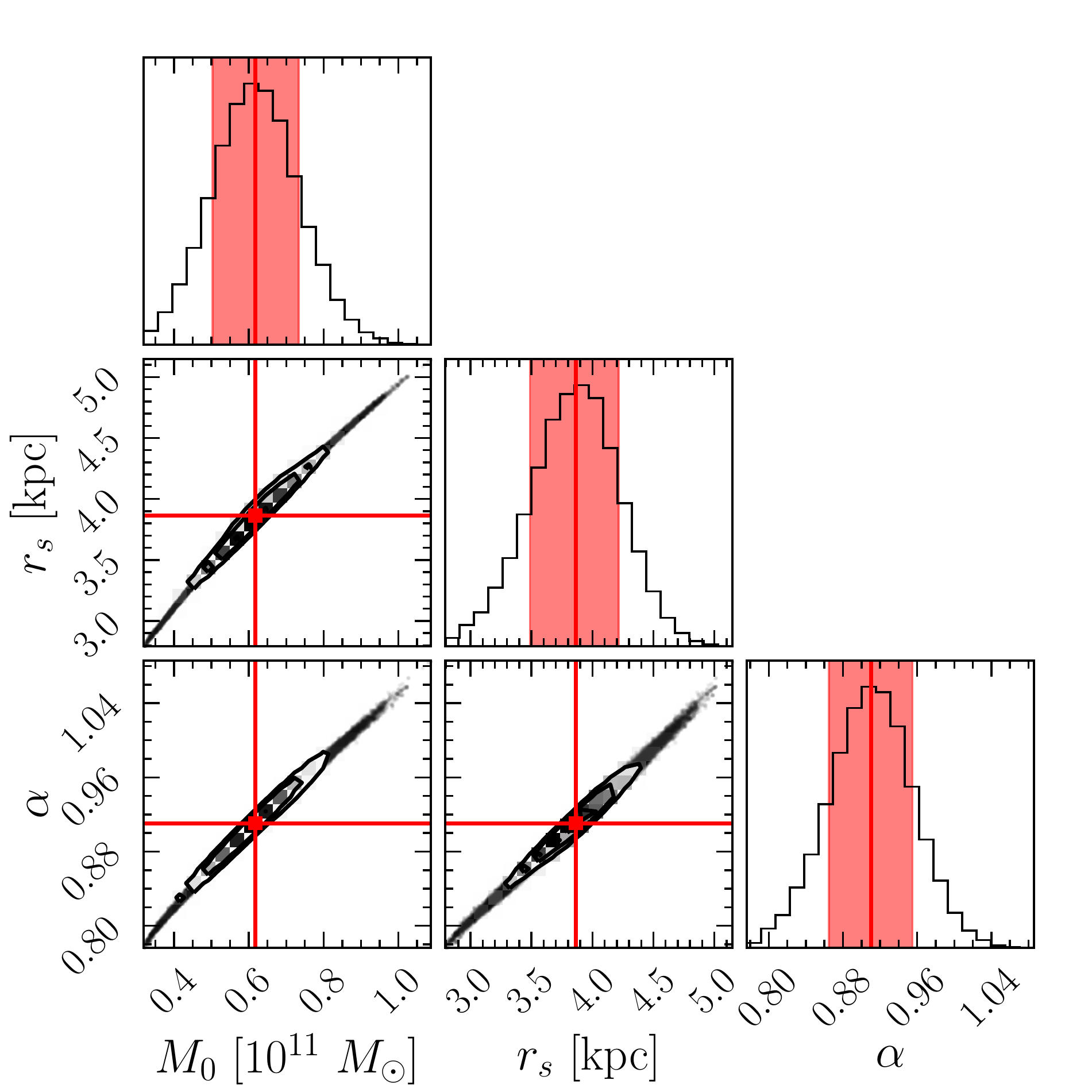}
    \includegraphics[width=0.45\textwidth]{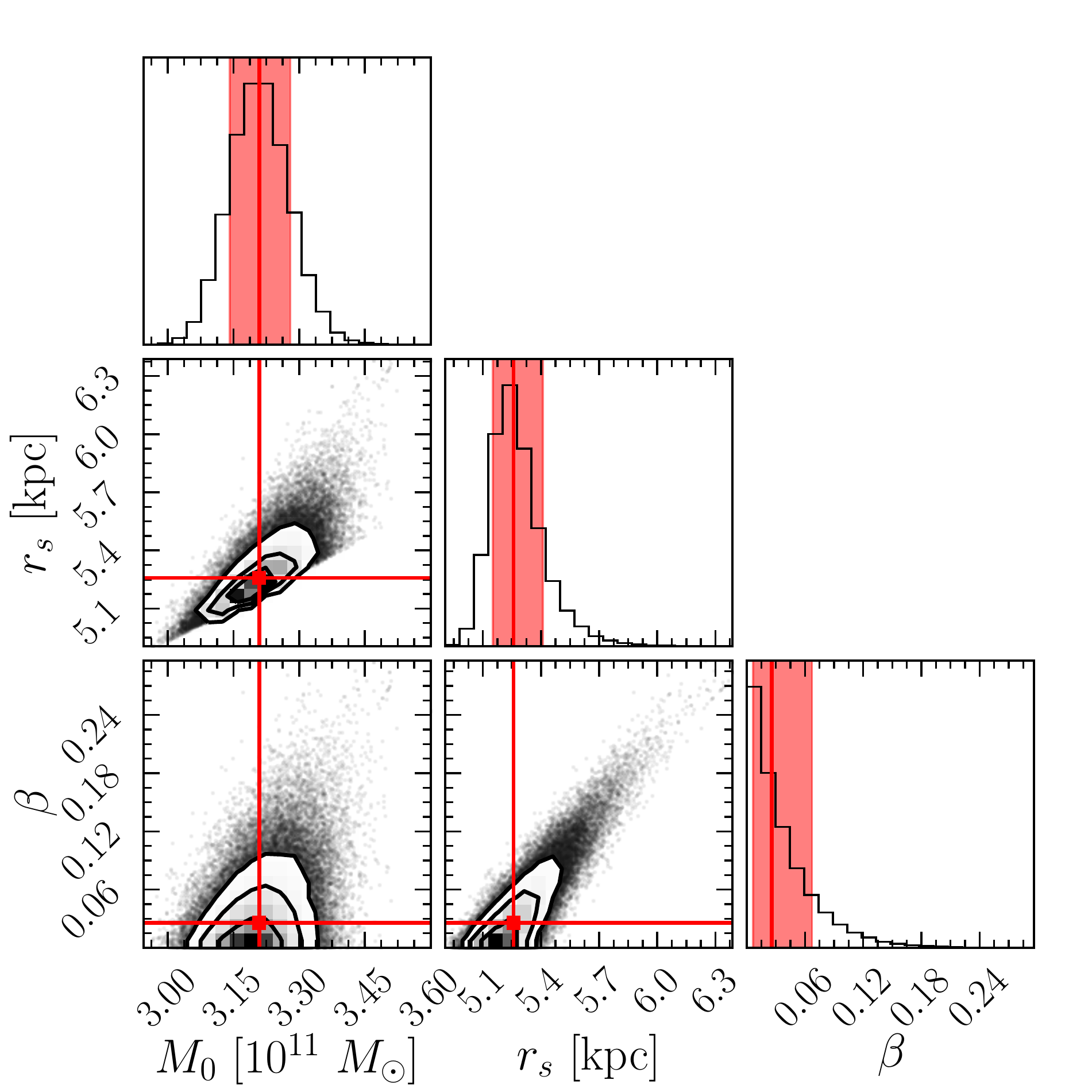}
    \caption{Posterior distribution of parameters for the Einasto (left) and gNFW (right) profile fit. The red line marks the median of the distribution, with the shaded region representing the 16$^{\rm{th}}$ to 84$^{\rm{th}}$ percentile. Note that the gNFW inner slope parameter $\beta$ posterior converges at $0$ (edge of the prior), suggesting a problematic fit as described in Section~\ref{sec:core_center}.}
    \label{fig:post_dtb}
\end{figure*}

The posterior distributions for both profile fits are shown in Figure~\ref{fig:post_dtb}. We take the median of the posterior distribution as the final fitted parameters for the DM profiles, listed in Table~\ref{tab:fit_res}. The statistical uncertainties are estimated with the 16$^{\rm{th}}$ and 84$^{\rm{th}}$ percentiles in the posterior distributions.
Virial masses ($M_{200}$) and radii ($r_{200}$), as well as concentration parameters ($c_{200}$), are calculated based on the best-fit parameters in the density profiles, assuming the cosmological parameters from \citet{planck20}. They are defined such that the average energy density within $r_{200}$ is $200$ times the critical density of the universe today, with $M_{200}$ the corresponding mass enclosed within this radius. $c_{200}$ is then defined as 
\begin{equation}
    c_{200} = \frac{r_{200}}{r_{-2}},
    \label{eq:c_200}
\end{equation}
where $r_{-2}$ is the radius at which the slope of the density profile $\frac{d \ln{\rho}}{d \ln{r}} = -2$. Given the definition of the profiles in Equation~\ref{eq:dens_gNFW} and~\ref{eq:dens_einasto}, $r_{-2} = (2-\beta)r_{s}$ for the gNFW profile and $r_{-2} = (2/\alpha)^{1/\alpha} r_{s}$ for the Einasto profile. We additionally compute quantities relevant to DM detection experiments, specifically the local DM density and $J$-factor.

\begin{figure*}
    \centering
    \includegraphics[width=\textwidth]{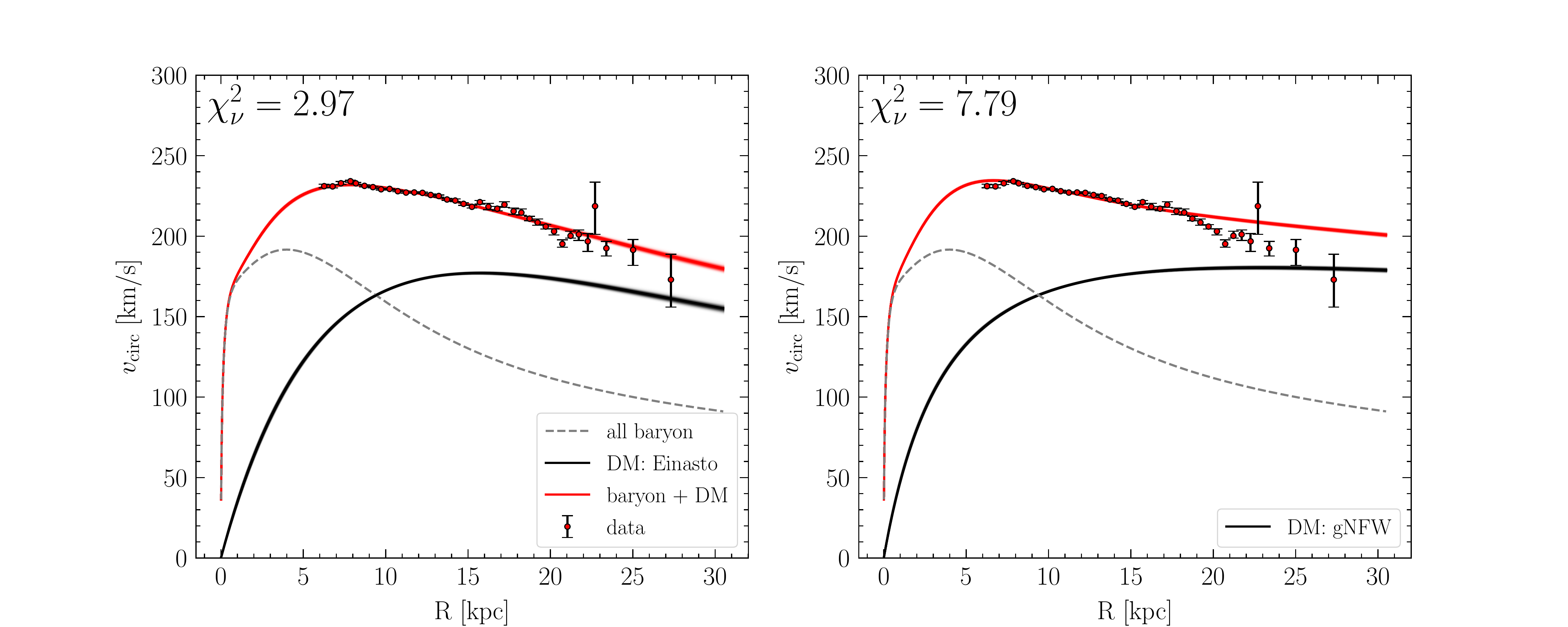}
    \caption{Comparison between Einasto(left) and gNFW(right) profile fit to our data. 
    NOTE: the Einasto is the best-fit profile; the gNFW profile fit is shown for comparison as discussed in Section~\ref{sec:result}.
    }
    \label{fig:halo_fit_compare}
\end{figure*}

The result suggests a highly cored DM profile in both models. Inspecting the two models further, we observe a clear distinction in the quality of the fit arising from the two profiles. We note that the best-fit model for the gNFW model has an inner power-law slope $\beta$ very close to 0, the edge of the prior. This is a clear sign of a problematic fit, as one can also see from the model circular velocity and the reduced $\chi^2$ values shown in Figure~\ref{fig:halo_fit_compare}. We report the final fitted values for the gNFW profile here in Table~\ref{tab:fit_res} merely for illustrative purposes and do not recommend using them for further analysis, given its reduced $\chi^2$ of \chisg. The Einasto profile, on the other hand, presents a much better fit to the data with a reduced $\chi^2$ of \chisE, as shown in the right panel of Figure~\ref{fig:halo_fit_compare}. 

\begin{table*}
\caption{\textsc{emcee} fitted results for Einasto and gNFW dark matter halo profiles. The uncertainties reported are purely statistical, as described in Section~\ref{sec:gNFW_einasto}. 
NOTE: the Einasto is the best-fit profile; the gNFW profile fit is shown for comparison as discussed in Section~\ref{sec:result}.
}
\label{tab:fit_res}
\begin{threeparttable}[t]
\centering
\begin{tabular}{lccc}
\hline 
 & 
Einasto & 
gNFW & 
Prior \\
\hline 

\textbf{Normalization Mass ($M_0$)} & \MnorE & \Mnorg & $[0.1,1000]\times10^{11}$~\msun\tnote{a}\\
\textbf{Scale Radius ($r_s$)} & \rscaE & \rscag & $[0,20]$~kpc \\
\textbf{Slope Parameter ($\alpha,\beta$)} & \slopE & \slopg & $[0,2]$ \\
\hline 
\textbf{Virial Mass ($M_{200}$)} & \MvirE & \Mvirg & -- \\
\textbf{Virial Radius ($r_{200}$)} & \rvirE & \rvirg & -- \\
\textbf{Concentration ($c_{200}$)} & \concE & \concg & -- \\
\textbf{Local Dark Matter Density ($\rho_{\rm{DM},\odot}$)} & \rhoSE & \rhoSg & -- \\
\textbf{$J$-factor ($J(\theta < 15^{\circ})$)} & \JfacE & \Jfacg & -- \\
\hline 
\textbf{$\chi^2$ per d.o.f. ($\chi_{\nu}^2$)} & \chisE & \chisg & -- \\
\hline

\end{tabular}
\begin{tablenotes}
     \item[a] Fitted in logarithmic scale.
\end{tablenotes}

\end{threeparttable}
\end{table*}

We note that the best-fit parameter uncertainties are purely statistical and thus the lower limits of any expected total uncertainties. 
The systematic uncertainties, as shown in Figure~\ref{fig:sys_err}, can reach over $15\%$ at $R > 22$\,kpc, significantly increasing the total uncertainties on the last few data points of the curve, which is expected from the lack of stars. 

A detailed study of the systematic uncertainties effect on the fit, along with concurrent fits of both the baryonic and DM components, will be shown in an upcoming work. 
However, we do not expect these uncertainties to result in a qualitative difference in the final results. 
We note that the systematic uncertainties are insufficient to explain the decline in the circular velocity curve starting at $R\sim 15$\,kpc, as shown in Figure~\ref{fig:vcirc_compare}. 
While the dominant systematic uncertainty from the neglected asymmetric drift correction term can potentially make the outermost circular velocity measurement to be consistent with $\sim220$\,\kmsec, we note that the decline is, in fact, established with measurements between $R=15$-$25$\,kpc, factoring in the systematic uncertainties.
Furthermore, as shown in \citet{desalas19}, factoring systematics uncertainties up to $12\%$ in the circular velocity measurements and varying baryonic model parameters primarily broaden the posterior distribution of the DM halo parameters. Additionally, in Sections~\ref{sec:core_formation} and \ref{sec:gNFW_einasto}, we show that our results are consistent with recent simulation studies. Also shown in Section~\ref{sec:core_formation}, they are also consistent with observational studies of the galactic centre DM profile.

\section{Discussion and Interpretation}
\label{sec:discussion}

This section is structured as follows.
We compare our circular velocity curve measurements to results from the literature in Section~\ref{sec:prev_lit}, where we find good agreements with other circular velocity curve studies.
We discuss the DM profile fit results in Sections~\ref{sec:core_center}, \ref{sec:core_formation} and \ref{sec:gNFW_einasto}, focusing on the origin of the cored centre and its implication on the formation history and predicted virial masses of the Milky Way.
We expand the discussion on the Milky Way mass to the context of the local group in Section~\ref{sec:big_picture} before moving on to Section~\ref{sec:direct_dm} and \ref{sec:indirect_dm}, where we discuss the best-fit Einasto profile in the context of DM direct and indirect detection experiments, respectively.

\subsection{Circular velocity curve comparison with previous work}
\label{sec:prev_lit}

\begin{figure*}
    \centering
    \includegraphics[width=\textwidth]{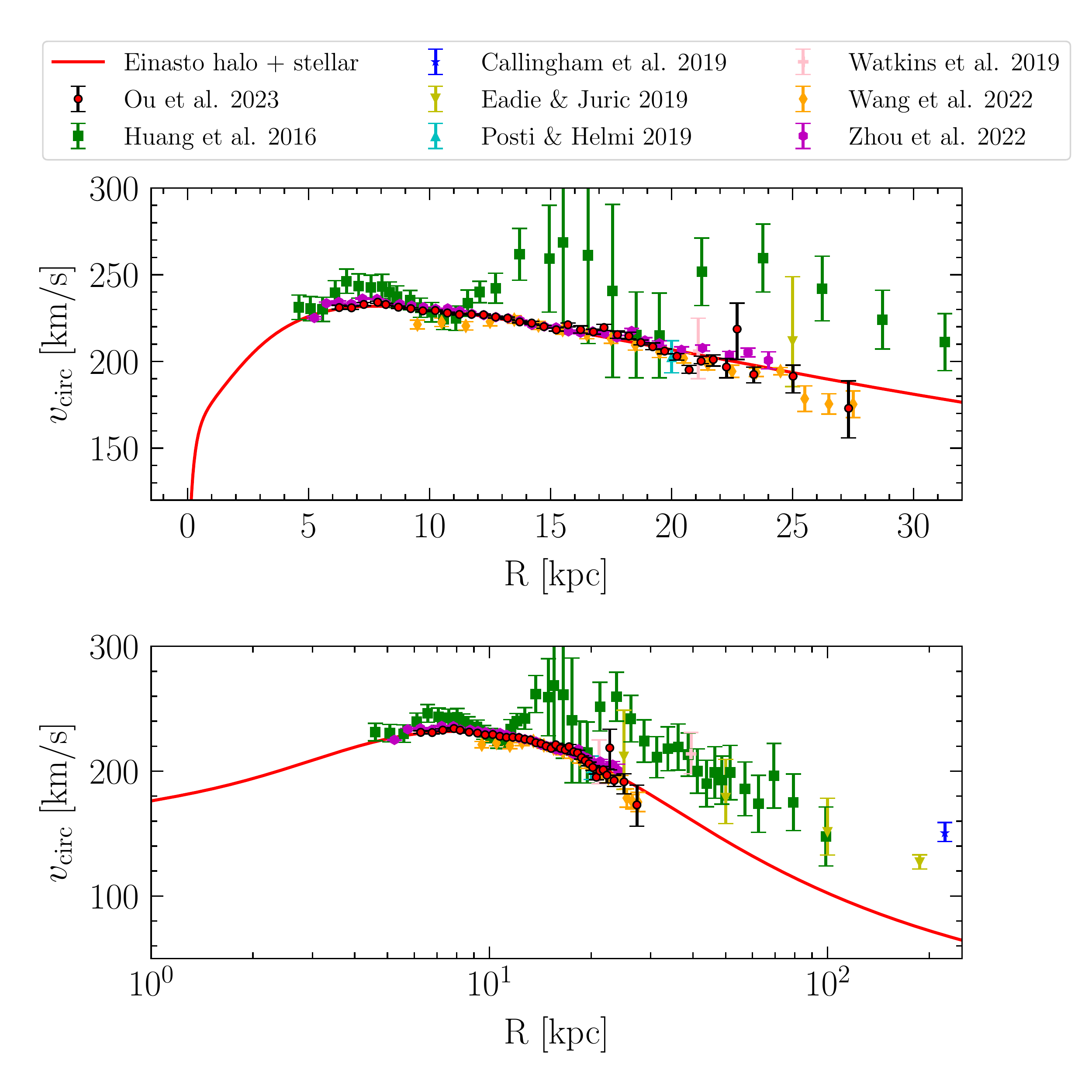}
    \caption{Best-fit model circular velocity curve (red curve) using the Einasto profile with circular velocity measurements from this study and previous literature values. The top(bottom) panel shows data for $R$ ranging from $\sim0$ to 30(110)\,kpc. Values are taken directly from the corresponding literature for \citet{huang16} and \citet{wang22}. For the virial/enclosed masses reported in \citet{callingham19}, \citet{eadie19} ,\citet{posti19a}, \citet{watkins19}, the circular velocities are calculated at virial/given radii.
    }
    \label{fig:einasto_w_lit}
\end{figure*}

Figure~\ref{fig:einasto_w_lit} shows a summary of our data and best-fit model circular velocity curve using the Einasto DM profile, combined with observational data from the literature. Our data and model show good consistency with those reported recently in \citet{wang22}, who also use the \Gaia\ DR3 parallaxes with a statistical deconvolution on the errors. In particular, \citet{wang22} also measures a significantly declining circular velocity at galactic radii greater than $25$\,kpc.

On the other hand, our results are systematically lower than those measured by \citet{huang16}, although statistically consistent within 2-sigma. As pointed out in \citet{eilers19}, we suspect such differences result from different tracer populations used for the analyses. 

At even larger galactic radii (lower panel of Figure~\ref{fig:einasto_w_lit}), our circular velocity curve model shows significant disagreement with circular velocities calculated from the Milky Way virial mass measurements \citep{callingham19,eadie19,posti19a,watkins19}. These studies use globular clusters and satellite galaxies around the Milky Way to estimate the enclosed mass of the Galaxy at large galactic radii (typically out to $R \sim 50-150$\,kpc) where individual stellar tracers are not available. We apply a conversion from the reported enclosed mass (or $M_{200}$) estimates in these studies to circular velocity estimates using the following relation
\begin{equation}
    v_{\rm{circ}} = \sqrt{\frac{GM}{R}},
    \label{eq:v_circ}
\end{equation}
where $G$ is the gravitational constant and $M$ is the enclosed mass within some radius $R$. In the case of $M_{200}$, we compute $R_{200}$ first assuming the cosmological parameter from \citet{planck20}. 

Our best-fit Einasto DM halo model predicts a circular velocity significantly lower than the values computed from the enclosed mass estimates. At $R \sim 110$\,kpc, the circular velocities converted from enclosed mass estimates from \citet{eadie19} and \citet{callingham19} are in the range of $130-150$\,\kmsec, whereas our best-fit model predicts $\sim 50$\,\kmsec\ at this distance from the galactic centre. Such disagreement is expected and in line with the fact that the best-fit parameters of our density profile predict a virial mass significantly lower than and inconsistent with those from any of the aforementioned studies. We further discuss these discrepancies and their potential causes in Section~\ref{sec:gNFW_einasto}.

\subsection{Cored centre: where does it come from?}
\label{sec:core_center}

We discuss the cause of the highly cored galactic centre in the Einasto and gNFW profile fits. As shown in Section~\ref{sec:result}, our measured circular velocity curve strongly prefers a core at the galactic centre. We argue that it is mainly driven by the curve's sharp decrease at $R > 20$\,kpc. 

While the fast decline in the circular velocity curve cannot be fully modelled by the gNFW profile, the posterior distribution of the inner power-law slope in Figure~\ref{fig:post_dtb} provides some hints as to why a cored centre is favoured. The problematic behaviour in the posterior distribution is expected when one considers the functional form of the gNFW profile. To begin with, the most well-measured data points at the inner galactic radius constrain the total mass enclosed up to $R \sim 20$\,kpc, where the drop in circular velocity begins. These data points thus apply a tight constraint on the normalization mass of the profile. As the galactic radius increases, the power law slope for the outer part of a gNFW profile is $(3-\beta)$, which, even at a maximum of 3, is not decreasing sharply enough to explain the drop in the data at $R > 20$\,kpc. Thus the posterior distribution for $\beta$ drifts towards 0, and the scale length prefers a smaller value than what has been reported in the literature to allow for the full $-3$ power-law decrease in density to start as early as possible. Consequently, this indicates that even though the gNFW profile is, by design, a bad model for our data, it is informative in pointing out the direction towards a different functional form that would allow both a core at the centre and a faster drop-off in density at larger radii as a function of galactocentric radius. 

The Einasto profile, therefore, presents itself as an ideal next choice, similarly preferring a core at the galactic centre. The exponential decrease of the profile allows for a much faster drop-off in density than the gNFW profile can at outer galactic radii. Compared to previous studies, we find a smaller scale radius and an $\alpha$ value $\sim 2$-$\sigma$ larger than 1. Again, these parameters suggest an extremely cored DM profile with faster than exponential drop off at galactic radii larger than $5.57$\,kpc.

It is thus essential to point out that the cored centre is partially driven by the functional form of the two profiles adopted in this study. The resulting best-fit profiles are principally only valid in the regime where there are circular velocity measurements available. Any inferences on the DM centre profile are built on the assumption that the functional form we choose to adopt is representative of the true profile. It is well possible that a more flexible profile could allow for a more cuspy centre and simultaneously predict circular velocities consistent with our measurements. A follow up study will focus on testing profiles with more flexible functional forms. For this study, we assume that the cored centre is physical and discuss the implications in the following sections.

\subsection{Dark matter core in the Milky Way}
\label{sec:core_formation}

While DM-only simulations generally predict cuspy DM profiles, recent studies have shown that cores can form in Milky Way-mass halos when incorporating baryonic feedback into simulations (Hydro simulations). \citet{lazar20}, using the Feedback In Realistic Environments (FIRE)-2 simulation, found that cores of sizes $0.5$-$2$\,kpc can be produced by feedback for galaxies with halo mass $\sim 10^{12}$~\msun. This phenomenon is consistent with previous studies using different simulation suites (e.g. FIRE-1 from \citealp{chan15} and NIHAO from \citealp{tollet16}). These hydrodynamic simulations show a consistent picture of core formation in galaxies evolving as a function of their stellar-to-halo mass ratio ($M_{\*}/M_{halo}$). The cores are found to be most significant (with core radii $\sim 1$-$5$\,kpc) for bright dwarf galaxies with $M_{\*}/M_{halo} \sim 5 \times 10^{-3}$ ($M_{\*} \sim 10^9$~\msun) (see, e.g. \citealt{lazar20}, Figure 7). As galaxies continue to grow and approach the size of the Milky Way (increasing $M_{\*}/M_{halo}$ to $\sim 10^{-1}$), the size of the DM core presents a significant scatter, indicating a complex picture of galaxies either retaining or destroying their cores. As pointed out by \citet{chan15}, whether the cores remain depends on the star formation episodes and DM accretion rates in the central regions of the galaxies. A survived core could indicate one or multiple starbursts after the central DM accretion has slowed down. A more detailed analysis is required to model if that is the case for the Milky Way, which is outside the scope of this study.

Evidence of a shallow cusp or core density profile for the Milky Way has also been found in previous dynamical studies of the galactic bulge. \citet{portail17} found that a power-law slope shallower than $-0.6$ is needed for the DM profile in the bulge region. They noted that an NFW profile could not simultaneously explain their best-fit baryonic mass distribution in the bulge and the flat rotation curve between 6-8\,kpc. Forcing both constraints requires the DM density to fall off more steeply than $-1$. At the same time, to avoid over-predicting the DM mass in the bulge, the slope must be shallower further in. This behaviour in the inner and outer slope is similar in spirit to what was described in Section~\ref{sec:core_center}. The fast decline in the circular velocity curve at outer galactic radii requires a steeper slope, whereas the flat (slowly declining) inner curve forces the slope to be shallow in the centre.

Our results provide further evidence of a core DM density profile for the Milky Way. By examining the Milky Way mass halos in the hydrodynamic simulations, we emphasize that the DM density profile of the Milky Way cannot be described by a universal picture but should be carefully considered in the context of galaxy formation.

\subsection{Virial masses: gNFW vs. Einasto}
\label{sec:gNFW_einasto}

We discuss the differences in the DM profile fitting result and its implication in this section. In particular, we focus on the different virial masses predicted by the two profiles. For the gNFW profile fit, we find the virial mass at \rvirg\ for the Milky Way DM halo at \Mvirg. For the Einasto profile fit, we find the virial mass at $r_{200}=$\rvirE\ for the Milky Way DM halo at \MvirE. Such drastic differences in virial mass estimates between the Einasto profile and the NFW/gNFW profile model have been studied in previous papers \citep{desalas19,jiao21}. \citet{desalas19} compared the results between a gNFW and Einasto profile fit using the same baryonic model (model B2 in the original study) of this study. They find that the Einasto profile consistently results in a lower virial mass estimate, with the gNFW profile giving $6.3^{+3.4}_{-1.3}\times10^{11}$~\msun\ and the Einasto profile giving $3.0^{+5.7}_{-1.2}\times10^{11}$~\msun. Our estimates, despite the poor fit in the gNFW case, are consistent with \citet{desalas19}. Similarly, \citet{jiao21} attempted to better account for the decline in the circular velocity curve found in \citet{eilers19} measurements. They found that the Milky Way virial mass could be as low as $2.60\times10^{11}$~\msun~with an Einasto profile fit, where the fit is mostly constrained by the last few data points from \citet{eilers19}. They concluded, however, that a higher virial mass up to $1.80\times10^{12}$~\msun~still could not be excluded when allowing the fit to be a little worse by forcing the gNFW profile. We see the same trend of the gNFW profile giving a systematically higher mass estimate than the Einasto profile. The data, especially at outer galactic radii ($R > 20$\,kpc), strongly prefer the Einasto profile. \citet{syloslabini23} combined the circular velocity curves from \citet{eilers19} and \citet{wang22} to study three DM models: an NFW halo model, a DM disc (DMD) model, and a model based on the Modified Newton Dynamics (MOND) theory. They find a similarly low virial mass with the NFW profile fit at $6.5^{+3}_{-3}\times10^{11}$~\msun, consistent with our gNFW result. The DMD model, on the other hand, predicts an even lower DM mass at $\sim 0.9\times10^{11}$~\msun.

The inconsistencies between results from disc stars and globular clusters/satellites as enclosed mass estimators likely come from systematics underlying the physical assumptions and different tracer populations. To estimate the enclosed mass from globular clusters or satellite tracers, it is often the case that some simplifying assumptions have to be made about the DM profile and that the final estimates are not sensitive to the inner structure of the halo. On the other hand, circular velocity curve measurements and modelling rely on stellar tracers and the assumption of an axisymmetric potential. The constraint on the density profile is only principally valid out to the innermost/outermost stars, and virial mass estimate is an extrapolation. Further analyses on the bound/unbound scenarios of individual satellites and globular clusters, given the density profile found in this study, are needed. We also call for more rigorous studies examining potential issues with assumptions that go into both methods for mapping the DM profile and mass. A more detailed and extensive analysis of systematics, such as various baryonic potential models, will be conducted in a future study.

\subsection{Milky Way mass in the big picture}
\label{sec:big_picture}

Aside from the inconsistency described above, it is perhaps not too surprising that the Milky Way is not as massive as thus far assumed. Specifically, while the mass estimates on the local group (LG) have been mostly steady at $\sim4-5\times10^{12}$~\msun\ (see, e.g. Table 4 of \citealt{chamberlain23}), the mass of M31 has been steadily trending up from $\sim 1.0\times10^{12}$~\msun\ to $\sim 2.5\times10^{12}$~\msun\ with new data and techniques becoming available (see, e.g. Figure 4 of \citealt{patel22}). While this does not indicate a Milky Way mass as low as what we find in this study, the general trend hints at a larger M31 to Milky Way mass ratio than previously used. If one makes a naive calculation by taking the lower LG mass estimates $3.4\times10^{12}$~\msun\ from \citet{benisty22} and the higher M31 mass estimates $2.99\times10^{12}$~\msun\ from \citet{patel22} in the \Gaia\ era, we see it is not entirely impossible that the Milky Way is an order of magnitude less massive than M31. We caution that this calculation is an oversimplification, not accounting for systematics concerning mass estimates from different methods, and should not be taken as a rigorous estimate of the Milky Way mass.

\subsection{Constraints on local DM density}
\label{sec:direct_dm}

Given the best-fit Einasto DM halo, the local DM density translates to \rhoSE. Despite the difference in the density profile model and the resulting virial mass, the local DM density we get is consistent with literature values, using either gNFW or Einasto profile. \citet{desalas19}, using the circular velocity curve measured by \citet{eilers19}, found $0.387^{+0.034}_{-0.036}$~GeV\,cm$^{-3}$ with a gNFW profile fit and $0.384^{+0.038}_{-0.034}$~GeV\,cm$^{-3}$ with an Einasto profile fit. Both are consistent with our results. 

This local DM density is also consistent with studies that exclusively use NFW/gNFW, with different input circular velocity curves and baryonic models. Global measurements from the pre-\Gaia\ era have been yielding values for the local DM density $\sim 0.4$~GeV\,cm$^{-3}$ (see e.g. \citealt[][Table 4]{read14}; \citealt{mcmillan17}). More recent studies such as \citet{zhou22} also find $0.39 \pm 0.03$~GeV\,cm$^{-3}$ with a gNFW profile combined with flexible disc scale radii, despite having systematically different circular velocity curve measurements. As discussed in \citet{mcmillan17}, the local DM density is not particularly sensitive to the inner power-law density slope of the profile. 

\subsection{Constraints on Indirect DM detections}
\label{sec:indirect_dm}

It is also interesting to consider the implication a highly cored Milky Way DM density profile has on the detectability of DM annihilation signals from the galactic centre. The $J$-factor, which incorporates the distribution of DM in an astrophysical system and determines the strength of the signal from annihilating DM, is essentially the square of the DM density integrated along the line-of-sight towards the galactic centre \citep{cirelli11}. For velocity-dependent DM annihilation, the factor is additionally dependent on the velocity distribution of the DM. This factor is crucial in modelling and translating the observed annihilation signal into the particle nature of DM (e.g. DM particle mass, annihilation cross-section, and Standard Model final states). Most studies on the Milky Way gamma-ray excess \citep{hooper11,daylan16} have been assuming an NFW or modified NFW profiles when carrying out the calculation of this factor, whether from an observing perspective \citep{ackermann17} or theory perspective \citep{boddy18}. A highly-cored Einasto profile will no doubt have a significant impact on the inferred J factor, and so far, studies are limited on this topic. \citet{hooper13} discussed the constraining power of potential DM annihilation signals from the galactic centre assuming different density profiles, including an Einasto and a constant core profile. They concluded that, even in the most conservative case, the galactic centre still provides a constraint on the dark matter annihilation cross-section as tight as those from dwarf galaxies with NFW profiles. 

Here we consider the simple case of self-annihilating DM signal as a result of our DM profile. We find the expected $J$-factor from a 15\degree\ view angle towards the galactic centre to be \JfacE\ for the best-fit Einasto profile. We additionally take the NFW profile parameters from \citet{cirelli11} to calculate the $J$-factor, which yields $\sim 118\times10^{22}$~GeV$^{2}$\,cm$^{-5}$. As expected, our $J$-factor value is significantly lower than that of the NFW profile. Even in the case of the poorly fit gNFW profile case, our $J$-factor estimate is $<25\%$ that of the NFW profile. Assuming some given DM particle property, our low $J$-factor estimates suggest an expected annihilating DM signal of only $10$-$20$\% of the current expected signal using the NFW profile. We do point out that the highly cored DM profile may provide new insights into the particle nature of DM particles, so a dedicated study is needed to ascertain the effect of our best-fit DM profile on the expected annihilation signal. Carrying out such a study is, unfortunately, outside the scope of this study.

\section{Conclusions}
\label{sec:conclusion}

In conclusion, we present the circular velocity curve for the Milky Way for $R \sim 6$-$27.5$\,kpc. We derive precise spectrophotometric parallaxes for $120,309$ luminous RGB stars using spectroscopic and photometric measurements. $33,335$ stars are selected as disc stars for circular velocity curve calculation using Jeans' equation. We extend the circular velocity curve beyond 25\,kpc with smaller statistical uncertainties compared to a previous study using a similar technique, thanks to a $\sim 50\%$ increase in sample size \citep{eilers19}. Our circular velocity curve shows good agreement with other recent studies that utilize \Gaia\ DR3 astrometry measurements. We find that the circular velocity curve declines at a faster rate at large galactic radii ($R > 20$\,kpc) compared to inner galactic radii. This trend was present, although not definitive, in \citet{eilers19} and is more clearly established in this study. 

We use the circular velocity curve to model the DM halo density profile, which is found to be likely cored. Two profiles, a gNFW profile and an Einasto profile, are fitted separately as the underlying DM profile for the Milky Way. We find that the Einasto profile presents a better fit to the data with the slope parameter $\alpha = $\slopE. The best-fit parameters for both profiles indicate a Milky Way DM halo with a core. We provide a simple intuitive explanation for the connection between the core and the shape of the circular velocity curve, namely that the core is a result of both the slowly declining inner and rapidly declining outer portions of the curve. We point out that a DM density core for a Milky Way-like galaxy can form principally in simulations \citep{lazar20}. The previous dynamic study of the galactic bulge by \citet{portail17} also shows evidence for a shallow cusp or core DM profile. The core may indicate a formation history with starbursts happening after the central DM accretion slowed down, but separate analyses combining Milky Way star formation history and accretion history are needed to fully understand this behaviour.

We discuss the implication of a cored Einasto profile on the virial mass estimates of the Milky Way. The predicted DM halo virial mass is \MvirE. While this value is overall lower than previous estimates, it is still consistent with recent studies which also use the circular velocity curve for virial mass estimates \citep{desalas19,jiao21,syloslabini23}. 

We stress that the cored profile and virial mass estimate are extrapolations from our measurements. Factoring in the potential systematics studied in this work, our circular velocity curve is only principally constraining between $\sim6$ to $25$\,kpc. We compare our results with mass estimates from Milky Way globular clusters and/or dwarf satellites dynamics \citep{eadie19,callingham19,correamagnus22}, as well as those from stellar streams \citep{vasiliev21,koposov23}. The discrepancy is most significant in regions outside of $R>30$\,kpc, where we do not directly probe.

The cored centre and virial mass results are thus derived by assuming a functional form for the underlying DM profile. In our case, our data prefers an Einasto profile over a gNFW profile. Future observations may help alleviate the need for an assumed functional form and directly bridge the gap between the circular velocity curve and satellite/stream results by providing stellar kinematics at overlapping $R$. Testing the different methodologies on simulations will also help identify potential systematic uncertainties between them.

In the context of DM detection experiments, we compute and discuss the local DM density and $J$-factor from our best-fit DM profile. On the one hand, we find local DM density to be \rhoSE, consistent with the literature. On the other hand, the $J$-factor (\JfacE) is found to be $\sim 13\%$ of that from a standard NFW profile, which is commonly used in galactic centre excess gamma-ray studies. 

Despite the potential systematic uncertainties, our study further demonstrates the power of constructing the circular velocity curve with the goal of probing the potential of the Galaxy. With large astrometric surveys such as \Gaia\ and a data-driven model, we are able to determine the circular velocity curve out to further distances for constraining the DM profile and formation history of the Milky Way. The results emphasize the uniqueness of the Milky Way DM halo and its potential implications on the nature of DM. This is a crucial step in eventually understanding the nature of DM and its role in galaxy formation in a cosmological context.

\section*{Software}

The analysis for this work was coded in \textsc{python} v. 3.7.6 \citep{python3} and includes its packages \textsc{IPython} \citep{perez07}, \textsc{numpy} \citep{vanderwalt11}, and \textsc{scipy} \citep{jones01}. We used \textsc{astropy} \citep{2013A&A...558A..33A,2018AJ....156..123A} and \textsc{ORIENT} \citep{mardini22} for coordinate transformation and orbit integration. We used \textsc{emcee} \citep{foremanmackey13} for fitting the DM profiles. Figures are generated with \textsc{matplotlib} \citep{hunter07}.

\section*{Acknowledgements}

We thank Ani~Chiti (University of Chicago), Nora~Shipp (MIT), Eugene~Vasiliev (University of Cambridge), and Pavel Mancera Pi\~{n}a (Leiden Observatory) for helpful discussions. A.F. acknowledges support from NSF grant AST-1716251.

This work presents results from the European Space Agency (ESA) space mission \emph{Gaia}. \Gaia\ data are being processed by the \Gaia\ Data Processing and Analysis Consortium (DPAC). Funding for the DPAC is provided by national institutions, in particular the institutions participating in the Gaia MultiLateral Agreement (MLA). The Gaia mission website is \url{https://www.cosmos.esa.int/gaia}. The Gaia archive website is \url{https://archives.esac.esa.int/gaia}.

This publication makes use of data products from the Two Micron All Sky Survey, which is a joint project of the University of Massachusetts and the Infrared Processing and Analysis Center/California Institute of Technology, funded by the National Aeronautics and Space Administration and the National Science Foundation.

This publication makes use of data products from the \textit{Widefield Infrared Survey Explorer}, which is a joint project of the University of California, Los Angeles, and the Jet Propulsion Laboratory/California Institute of Technology, funded by the National Aeronautics and Space Administration.

Funding for the Sloan Digital Sky Survey IV has been provided by the Alfred P. Sloan Foundation, the U.S. Department of Energy Office of Science, and the Participating Institutions. SDSS-IV acknowledges support and resources from the Center for High Performance Computing at the University of Utah. The SDSS website is \url{www.sdss4.org}.


This work used Stampede-2 under allocation number TG-PHY210118, part of the Extreme Science and Engineering Discovery Environment (XSEDE), which is supported by National Science Foundation grant number ACI-1548562. This work used the Engaging cluster supported by the Massachusetts Institute of Technology.

This research has made use of NASA's Astrophysics Data System Bibliographic Services; the arXiv pre-print server operated by Cornell University; the SIMBAD and VizieR databases hosted by the Strasbourg Astronomical Data Center.

\section*{Data Availability}

Spectrophotometric parallaxes of stars derived from this study are available for download as online material accompanying this work. 



\bibliographystyle{mnras}
\bibliography{xou} 



\appendix

\section{Potential non-disc contaminants}
\label{sec:contaminant}

\begin{figure*}
    \centering
    \includegraphics[width=\textwidth]{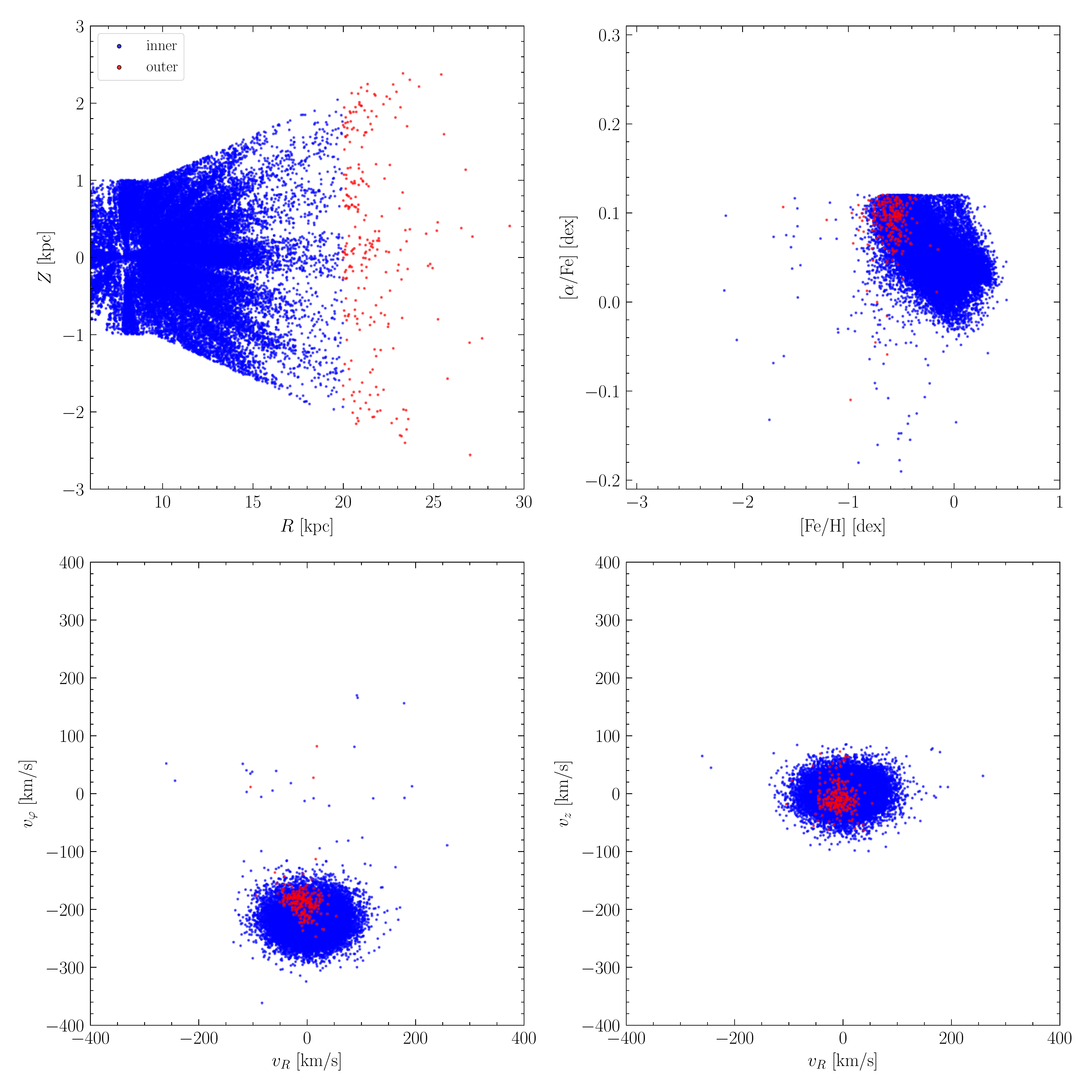}
    \caption{
    Comparison between stars with $R<20$\,kpc (blue) and stars with $R>20$\,kpc (red). The top-left panel shows the stars in cylindrical galactocentric coordinates. The top-right panel includes the $\alpha$-element abundances ([$\alpha$/Fe]) as a function of metallicity ([Fe/H]) with value taken from \textit{APOGEE}. The bottom two panels are the cylindrical velocities of the two samples.
    }
    \label{fig:comp_in_out}
\end{figure*}

We briefly discuss potential contaminants in our sample that could potentially bias the analysis. Figure~\ref{fig:comp_in_out} shows the sample divided at $R=20$\,kpc in chemodynamic spaces with $33,091$ stars for the ``inner'' sample and $244$ stars for the ``outer'' sample. We find no significant halo star contamination in the outer disc sample based on the space velocities. Compared to the inner sample, the metallicity of the outer sample is systematically lower on average than that of the inner sample. This is expected due to the radial metallicity gradient in the disc \citep{majewski17,gaia22}. Similarly, we expect the outer disc to have higher [$\alpha$/Fe] relative to the inner sample. 

We do note that there exist a few inner and outer stars with potentially high eccentricities (high $v_{R}$ and/or low $v_{\varphi}$). Removing them from the respective samples, however, make minimal differences to the derived circular velocity curve. For simplicity, and the fact that our calculation is carried out on $v_{R}$ and $v_{\varphi}$, we chose to not introduce any additional arbitrary manual cut on any of the velocities. Hence, we kept the high eccentricity stars, but we emphasize again that the result is qualitatively the same without these stars. 

\begin{figure*}
    \centering
    \includegraphics[width=0.9\textwidth]{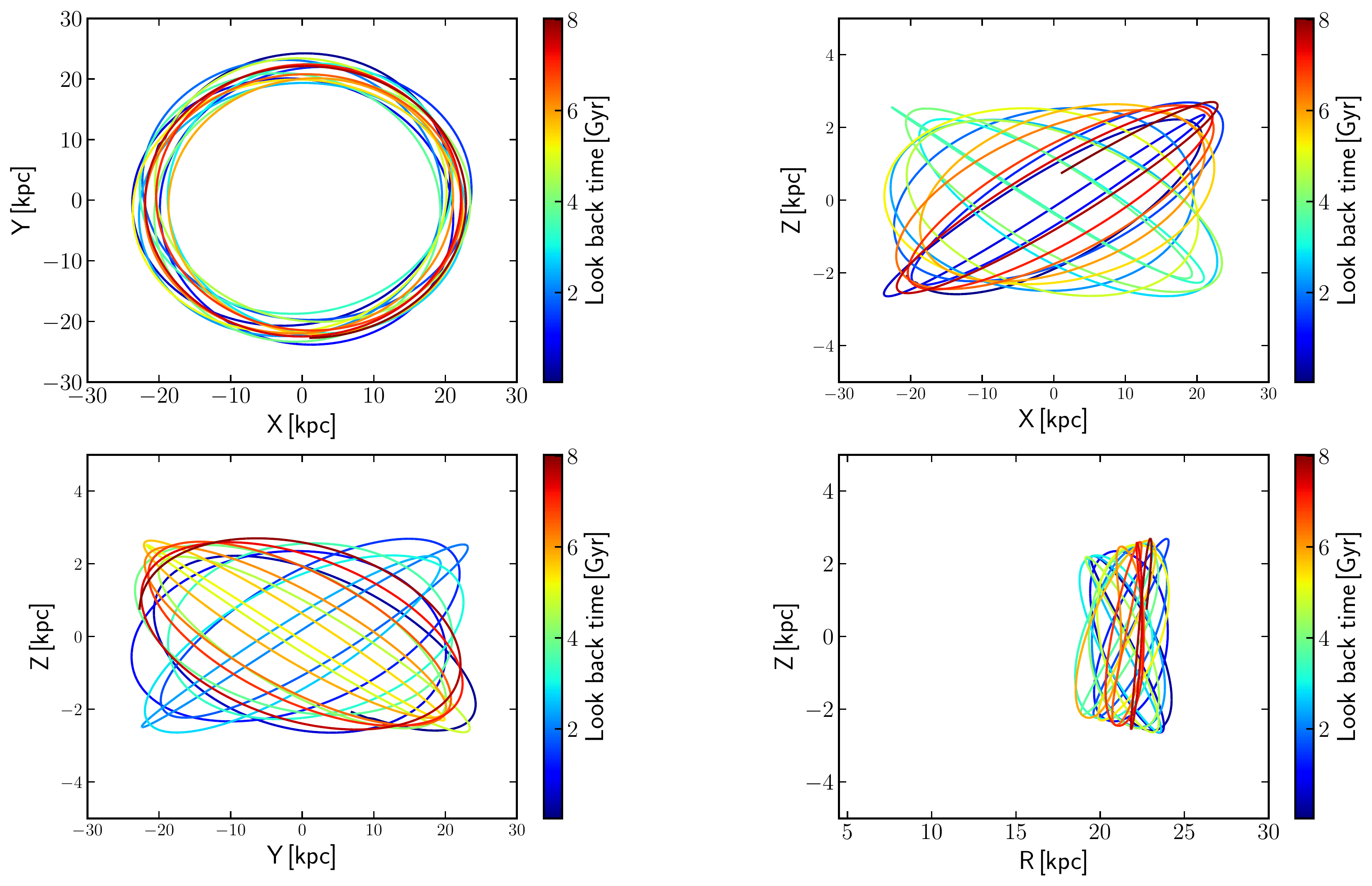}
    \caption{
    Projections of an outer disc star, \Gaia~DR3~244707655774974464, with a disc-like orbit in the X-Y (top-left), X-Z (top-right), Y-Z (bottom-left), and R-Z (bottom-right) planes. The orbit is color-coded by the look back time. The majority of the outer sample exhibits these very similar orbits.
    }
    \label{fig:orbit_disc}
\end{figure*}

\begin{figure*}
    \centering
    \includegraphics[width=0.9\textwidth]{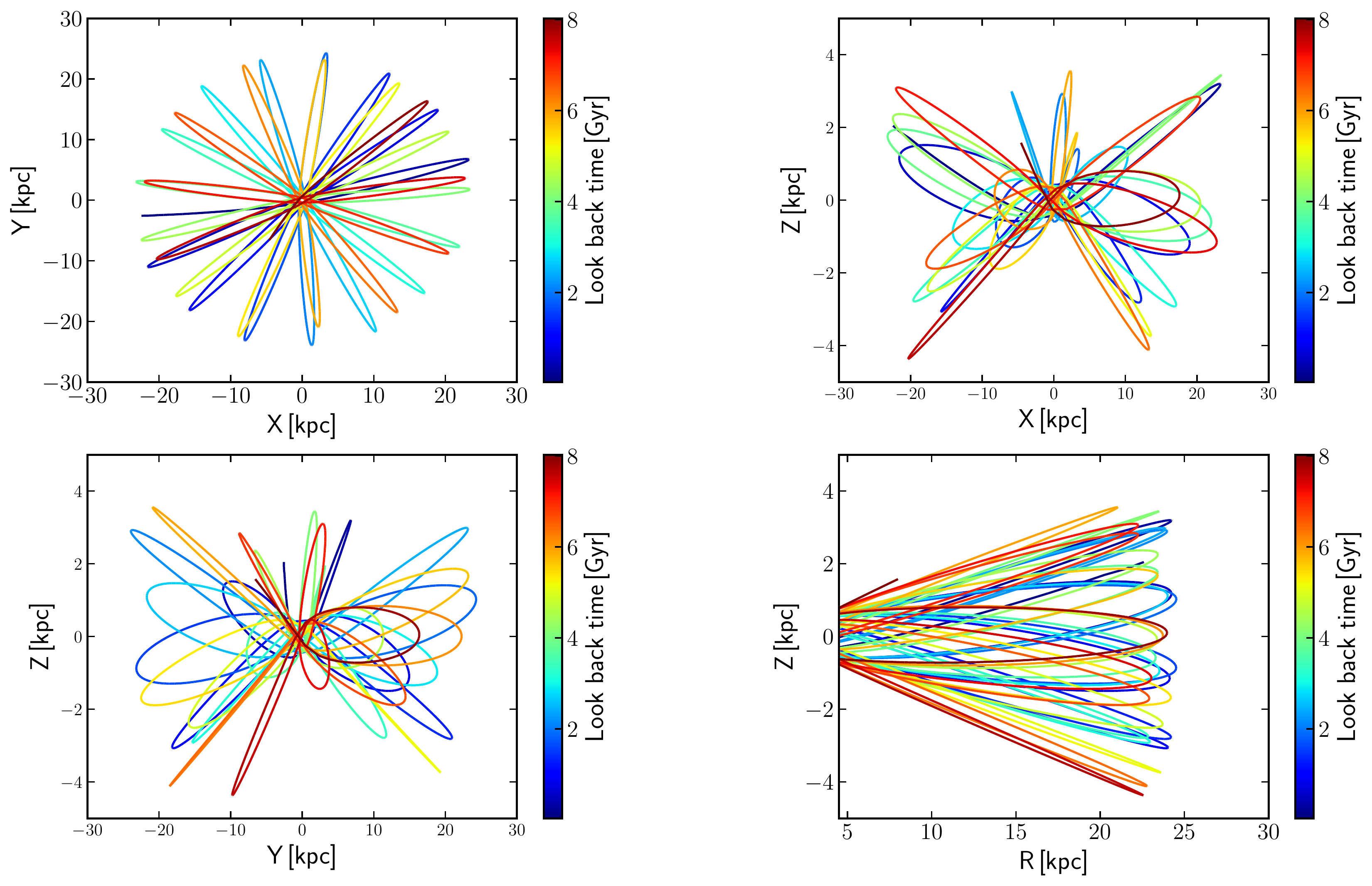}
    \caption{
    Projections of an outer disc star, \Gaia~DR3~3382839632449159040, with a highly eccentric orbit in the X-Y (top-left), X-Z (top-right), Y-Z (bottom-left), and R-Z (bottom-right) planes. The orbit is color-coded by the look back time. Only three out of the $244$ outer sample stars present such an orbit.
    }
    \label{fig:orbit_ecc}
\end{figure*}

We additionally perform orbital integrations with the \textsc{ORIENT} package \citep{mardini22} to examine the orbits of all outer disc stars. Most stars exhibit orbits that are near circular, as expected for a disc star. An example is shown in Figure~\ref{fig:orbit_disc}. Three stars with low $v_{\varphi}$ do present highly eccentric orbits, as shown with an example in Figure~\ref{fig:orbit_ecc}.

\bsp	
\label{lastpage}
\end{document}